\documentclass[11pt]{article}
\batchmode

\usepackage{epsfig}
\usepackage{amsfonts}
\usepackage{amsmath}
\usepackage{amssymb}
\usepackage{dsfont}
\usepackage{hyperref}
\usepackage{color}
\usepackage{cite}
\usepackage{braket}	
\usepackage{caption}
\usepackage{subcaption}	

\usepackage{feynmp-auto}

\RequirePackage[colorinlistoftodos,prependcaption,textsize=tiny]{todonotes}
\usepackage{soul}

\newcommand{\mpi}{m_{\pi}}

\newcommand{\dslash}[1]{#1 \llap{/\kern-0.5pt}}
\newcommand{\Dslash}[1]{#1 \llap{/\kern+1.5pt}}
\newcommand{\DDslash}[1]{#1 \llap{/\kern+2.3pt}}
\newcommand{\dslashh}[1]{#1 \llap{/\kern+1pt}}
\newcommand{\abs}[1]{\left|#1\right|}

\newcommand{\boldtau}{\mbox{\boldmath $\tau$}}
\newcommand{\boldpi}{\mbox{\boldmath $\pi$}}

\newcommand{\bea}{\begin{eqnarray}}
\newcommand{\eea}{\end{eqnarray}}
\newcommand{\bma}{\begin{pmatrix}}
\newcommand{\ema}{\end{pmatrix}}

\renewcommand{\vec}[1]{\boldsymbol{#1}}
\newcommand{\uvec}[1]{\hat{\vec{#1}}}

\newcommand{\ClebschGordan}[3]{
  \left\langle \, #1 , \,  #2 \, \middle\vert \, #3 \, \right\rangle
}
\newcommand{\CG}[6]{\ClebschGordan{ {#1}\,{#4} }{ {#2}\,{#5} }{ {#3}\,{#6} } }

\newcommand{\diff}[1]{\text{d}{#1} \, }
\newcommand{\MeV}{\mbox{ MeV}}
\newcommand{\fm}{\mbox{ fm}}

\newcommand{\MyIncFig}[4]
{%
\par
\begin{figure}[htbp]
	\centering
	\includegraphics*[#2]{#1}
	\caption{\label{#3}#4}
\end{figure}%
\par
}

\begin{document}
\errorstopmode
\begin{titlepage}
\hypersetup{pageanchor=false}

\begin{flushright}
Nikhef 2017-022
\end{flushright}

\vspace{1.0cm}

\begin{center}
{\LARGE  \bf 
First-principle calculations of  \\[0.3cm] 
Dark Matter scattering off light nuclei
}
\vspace{1.4cm}

{\large \bf  C. K{\"o}rber$^{a}$, A. Nogga$^{a,b}$, J. de Vries$^c$ }
\vspace{0.5cm}

{\large $^a$ {\it Institute for Advanced Simulations 4, Institute f{\"u}r Kernphysik 3, J{\"u}lich Center for Hadron Physics, 
and JARA - High Performance Computing, Forschungszentrum J{\"u}lich, D-52425 J{\"u}lich, Germany}\\}%
\vspace{.25cm}
{\large $^b$ {\it Department of Physics and Astronomy, Ohio University, Athens, OH 45701, USA}\\}%
\vspace{.25cm}
{\large $^c$ {\it Nikhef, Theory Group, Science Park 105, 1098 XG, Amsterdam, The Netherlands}\\}%
\vspace{.25cm}
\end{center}

\vspace{.5cm}

\begin{abstract}
We study the scattering of Dark Matter particles off various light nuclei within the framework of chiral effective field theory. 
We focus on scalar interactions and include one- and two-nucleon scattering processes whose form and strength are dictated by chiral symmetry.
The nuclear wave functions are calculated from chiral effective field theory interactions as well and we investigate the convergence pattern of the chiral expansion in the nuclear potential and the Dark Matter-nucleus currents. This allows us to provide a systematic uncertainty estimate of our calculations. We provide results for ${}^2$H, ${}^3$H, and ${}^3$He nuclei which are theoretically interesting and the latter is a potential target for experiments. We show that two-nucleon currents can be systematically included but are generally smaller than predicted by power counting and suffer from significant theoretical uncertainties even in light nuclei. We demonstrate that accurate high-order wave functions are necessary in order to incorporate two-nucleon currents. We discuss scenarios in which one-nucleon contributions are suppressed such that higher-order currents become dominant.

\end{abstract}

\vfill
\end{titlepage}
\hypersetup{pageanchor=true}

\section{Introduction}

Despite the lack of a conclusive signal in a direct, indirect, or accelerator experiment, the astrophysical and cosmological evidence for Dark Matter (DM) remains very strong \cite{Bertone:2016nfn}. Direct detection experiments have greatly improved in recent decades \cite{Akerib:2016vxi, Fu:2016ega} and are expected to improve even further in the near future \cite{Aprile:2015uzo, Akerib:2015cja}. These experiments have the great potential to not only discover the existence of DM but also to unravel its nature. The interpretation of a signal (or lack thereof) however, requires input from hadronic and nuclear physics in order to connect the experimental results to more fundamental interactions between DM and Standard Model (SM) particles. 

The nuclear physics aspects of DM direct detection has recently been discussed in several effective field theory (EFT) frameworks \cite{Fan:2010gt, Fitzpatrick:2012ix,Cirigliano:2012pq,Klos:2013rwa,Hill:2014yxa,Hoferichter:2015ipa , Bishara:2016hek}. For instance, Ref.~\cite{Fitzpatrick:2012ix} constructed an EFT description at the DM-nucleon level by constructing all possible interactions up to a certain order in momentum transfer and performed shell-model calculations of the associated nuclear responses for different target nuclei. A different approach includes the consequences of spontaneous breaking of chiral symmetry and the associated pions as pseudo-Goldstone bosons. Starting from a given set of interactions between DM and SM fields (in particular light quarks and gluons), this chiral-EFT approach allows for the systematic derivation of DM-hadron interactions. In this approach, DM-nucleus scattering does not solely arise from DM-nucleon interactions but also from two-nucleon currents arising from DM-pion interactions \cite{Prezeau:2003sv,Cirigliano:2012pq,Klos:2013rwa,Hoferichter:2015ipa}. The chiral-EFT power counting predicts a hierarchy of different terms which can be systematically included. For example, Refs.~\cite{Klos:2013rwa,Hoferichter:2016nvd} investigated in detail several two-nucleon interactions and performed shell-model calculations of the associated structure factors.

The effects of multi-nucleon interactions are particularly interesting as these can potentially lead to a very distinct  dependence of the DM-nucleus cross section on $A$ (the atomic number) and $Z$ (number of protons) of the target nucleus with respect to the standard DM-nucleon contributions. For instance, scalar DM-quark or DM-gluon interactions lead to similar DM-nucleon interactions but quite distinct two-nucleon interactions. If two-nucleon contributions can be probed by experiments on different target nuclei it might therefore be possible to unravel different DM-quark/gluon interactions even if they both lead to spin-independent scattering processes. Two-nucleon corrections for spin-independent scattering were studied in several previous works \cite{Prezeau:2003sv,Cirigliano:2012pq,Klos:2013rwa,Hoferichter:2015ipa,Beane:2013kca}. In general, the corrections are found to be relatively small. Ref.~\cite{Hoferichter:2016nvd} found $\mathcal O(10\%)$ corrections for ${}^{132}$Xe and somewhat smaller for light nuclei \cite{Beane:2013kca}, but they can become much more important in specific scenarios where the one-body contributions are suppressed \cite{Prezeau:2003sv,Cirigliano:2012pq,Cirigliano:2013zta}. 
 
Clearly, the above statements depend crucially on the accuracy of the nuclear calculations, which we investigate in this work. We also apply the chiral-EFT framework for DM scattering, but we wish to simultaneously describe the nuclear wave functions within the same chiral-EFT approach. This allows for a first-principle calculation, \textit{i.e.} starting from an assumed (set of) DM-SM interaction(s), of the DM-nucleus cross sections. The complexities of many-body nuclear physics limits this approach to light nuclei and here we perform calculations of DM scattering of the first few bound nuclei. Such nuclei are interesting for both theoretical and experimental reasons. Theoretically, light nuclei are great testing laboraties as they can be described from first principles to high accuracies. On the other hand, for example, helium isotopes are potential experimental targets \cite{Guo:2013dt, SNONEWS, Schutz:2016tid} as they are sensitive to relatively light DM (below 10 GeV) \cite{Profumo:2015oya}  and they can potentially be used for directional detection purposes \cite{Franarin:2016ppr,Gazda:2016mrp}. Our calculations provide direct input for the interpretation of these experiments. Very recently, Ref.~\cite{Gazda:2016mrp} performed an analysis of DM scattering of light nuclei in similar spirit to our work but extended to the ${}^4$He case. The main difference in the applied method with respect to our work is that Ref.~\cite{Gazda:2016mrp} applies chiral wave functions at fixed chiral order (next-to-next-to-leading order) obtained from the no-core shell model, in combination with one-nucleon currents not derived from chiral perturbation theory but taken from Ref.~\cite{Fitzpatrick:2012ix}.

In this work, we wish to investigate several important questions: how dependent are DM scattering cross sections on the nuclear potential and the resulting nuclear wave functions? That is, how large are the intrinsic nuclear uncertainties? How important are two-nucleon operators compared to standard one-body interactions and, crucially, how accurately can we calculate such contributions? 
As mentioned, we focus on scattering off the deuteron (${}^2$H), triton (${}^3$H), and the helium isotope ${}^3$He, while results for ${}^4$He are presented in upcoming work. These are nuclei for which we can solve the bound-state equations with direct methods for a given nuclear potential (including three-nucleon forces).  We use these nuclei as theoretical laboratories and aim to extend our framework to heavier nuclei with more sophisticated many-body techniques.
Tremendous progress has been made in recent years in first-principle calculations of medium-heavy nuclei by use of nuclear quantum Monte Carlo methods \cite{Carlson:2014vla} and nuclear lattice-EFT \cite{Lahde:2013uqa}. Anticipating similar progress for DM direct detection, our calculations perform a benchmark for such future studies. 

This paper is organized as follows. In Sect.~\ref{sec:framework} we introduce the DM interactions we focus on and introduce the chiral EFT framework. We derive the one- and two-nucleon currents and discuss the power counting. In Sect.~\ref{sec:waves} we describe the calculation of the nuclear wave functions and the scattering cross section. We explain how we estimate the theoretical uncertainties associated to our calculations. We present our main results in Sect.~\ref{sec:results}, where we also discuss scenarios where leading-order cross sections are suppressed and subleading effects become dominant. Finally, we summarize and give an outlook for future work in Sect.~\ref{Summary}. 

\section{Scalar Dark Matter currents and the chiral effective field theory framework}\label{sec:framework}
In this work we focus on scalar interactions between DM and light quarks and gluons
\begin{equation}\label{Eq1}
\mathcal L_{\chi} = \bar \chi \chi \left (  c_u\,m_u\,\bar u u + c_d\,m_d\,\bar d d+ c_s\,m_s\,\bar s s + c_G \,\alpha_s G_{\mu\nu}^a G^{\mu\nu\,a} \right )\ ,
\end{equation}
in terms of a DM spin-$1/2$ Dirac fermion $\chi$ (the cases of scalar, vector, or Majorana DM particles are almost identical for the scalar interactions under consideration), light quark fields $u$, D, S, the gluon field strength $G_{\mu\nu}^a$, $\alpha_s$ the strong coupling, and three coupling constants $c_{u,d,s,G}$ of mass dimension $(-3)$ that parametrize the strength of the DM-SM interactions. For convenience, we have assumed the DM-quark operators to scale with the quark masses, $m_{u,d,s}$.

The Lagrangian in Eq.~\eqref{Eq1} is taken at a scale around $1$ GeV and at lower energies we match to effective DM-hadron interactions. While we focus on scalar interactions as these are well motivated, for instance via Higgs-portal DM, and have interesting two-nucleon currents \cite{Prezeau:2003sv,Cirigliano:2012pq,Hoferichter:2016nvd},  clearly there can be other interactions, \textit{e.g.} vector, axial-vector, or tensor interactions, or DM-photon couplings, and we leave the more general case to a forthcoming publication. The couplings $c_{u,d,s,G}$ include contributions from potential couplings to heavier quarks which have been integrated out \cite{Shifman:1978zn}. 

We wish to describe the scattering process between DM and light nuclei completely within the framework of chiral EFT. Our starting point is the QCD Lagrangian supplemented by the DM interactions in Eq.~\eqref{Eq1}.
By constructing the most general Lagrangian that incorporates the symmetries of the microscopic theory (QCD supplemented with DM interactions) and their explicit and spontaneous breaking, in terms of the relevant low-energy degrees of freedom (DM fields, pions, nucleons, and, in principle, heavier hadrons), we obtain chiral perturbation theory ($\chi$PT), which is the low-energy equivalent of QCD. The power of $\chi$PT is that it can be used to calculate hadronic and nuclear observables in perturbation theory where $p/\Lambda_\chi$ is the expansion parameter in terms of P, the low-momentum scale of the process, and $\Lambda_\chi \simeq 1$ GeV the chiral-symmetry-breaking scale. 

While $\chi$PT allows for the derivation of the form of the interactions, each term is associated with a low-energy constant (LEC) that captures the non-perturbative nature of low-energy QCD. These LECs need to be fitted to data or calculated with non-perturbative methods such as lattice QCD. $\chi$PT has been extended to the multi-nucleon sector where it is usually called chiral EFT ($\chi$EFT) and this has led to the derivation of the strong nucleon-nucleon ($N\!N$) potential up to next-to-next-to-next-to-next-to-leading  order (N${}^4$LO)\cite{Epelbaum:2014sza}. Throughout this work we apply $SU(2)$ $\chi$PT instead of $SU(3)$ $\chi$PT as the extension of the latter theory to include nuclear forces has not been developed to the same accuracy as for the $SU(2)$ case. We therefore do not include dynamical effects of the strange pseudo-Goldstone bosons: kaon- and eta-mesons. 

Pions play an important role in $\chi$PT as they emerge as Goldstone bosons of the spontaneously broken chiral symmetry of QCD to the subgroup of isospin $SU(2)_L\times SU(2)_R\rightarrow SU(2)_I$.  Because chiral symmetry is only an approximate symmetry, being violated by quark masses and charges and, in our case, the DM-quark scalar interactions,  the pions obtain a small mass $m_\pi^2 \sim m_q$. The smallness of the symmetry-breaking terms fortunately ensures that the chiral-symmetry-breaking sources can be incorporated in the chiral expansion. The $\chi$EFT Lagrangian is then obtained by adding to all chiral-invariant interactions, all interactions that break chiral symmetry in the same way as the quark level chiral-symmetry-breaking sources. The infinite numbers of terms can be ordered by the chiral index $\Delta = d + n/2-2$, where D counts the number of derivatives and quark mass insertions (a quark mass insertion increases D by $2$ because $m_q\sim\mpi^2\sim p^2$) and $n$ the number of nucleon fields \cite{Weinberg:1978kz,Gasser:1983yg}.  Since we are interested in processes where a single DM particle scatters off a nucleus we will only consider chiral interactions that are linear in the couplings $c_{u,d,s,G}$. In what follows we introduce $Q= p/\Lambda_\chi$ as the expansion parameter.

To obtain the nuclear wave function we require the strong nucleon-nucleon potential. At leading order (LO) $(\mathcal O(Q^0))$ the potential consists of a one-pion-exchange (OPE) diagram and two short-range nucleon-nucleon interactions \cite{Weinberg:1990rz}. At next-to-leading order (NLO) $(\mathcal O(Q^2))$ one finds corrections to the OPE  diagrams, several two-pion-exchange diagrams, and subleading contact interactions \cite{Ordonez:1993tn}. At N${}^2$LO $(\mathcal O(Q^3))$ additional TPE diagrams appear that arise from $\pi\pi$-nucleon interactions with chiral index $\Delta=1$, the $c_i$ interactions \cite{Bernard:1995dp}, which also give rise to three-nucleon forces. The number of terms grows at even higher order \cite{Epelbaum:2014sza}, although how many terms are relevant depends on the process under investigation. 

The LECs appearing in the potential are fitted to pion-nucleon scattering data (see e.g. Ref.~\cite{Hoferichter:2015tha})  and the few-nucleon database and then other nuclear observables can be predicted. The scattering and bound-state equations are typically divergent and a coordinate-space cut-off is applied to regulate the integrals. Of course, observables should not depend on the chosen cut-off, but in numerical calculations explicit cut-off independence is lost. The LECs appearing in the nucleon-nucleon potential are fitted (at each order) for different values of the applied cut-off. By varying the chiral order of the potential and the cut-off we can test both the chiral convergence and the cut-off dependence of our results, allowing for a well-defined uncertainty estimate. We provide more details of this procedure below.

\subsection{Currents for isoscalar and isovector DM-quark interactions}\label{PC}

The second part of the calculation involves the chiral expansion of DM-hadron interactions. For the scalar interactions under consideration this has been studied in detail in, for example, Refs.~\cite{ Cirigliano:2012pq,Hoferichter:2016nvd,Bishara:2016hek}. Here we repeat the analysis for completeness and add a few comments about higher-order corrections. We begin by considering scalar interactions involving up and down quarks. These can be treated in $\chi$PT as ordinary quark mass terms by replacing the usual spurion field $\chi$ 
\begin{equation}
\chi = 2 B \mathcal M \rightarrow 2 B \left[ \mathcal M - \mathrm{diag}(m_u c_u\,\bar \chi \chi,\,m_d c_d\,\bar \chi \chi) \right]\, ,
\end{equation}
where $\mathcal M = \mathrm{diag}(m_u,\,m_d)$ is the quark mass matrix. The leading terms in the DM chiral Lagrangian are then given by
\begin{eqnarray}\label{eq:2.3}
\mathcal L_{\chi,q} &=& \frac{f_\pi^2}{4} \textrm{Tr}\, [U^{\dagger} \chi + U \chi^{\dagger}] +  c_1  \textrm{Tr}(\chi_+) \bar N N + c_5\bar N \hat{\chi}_+   N\ ,
\end{eqnarray}
where $N = (p\, , \, n)^T$ is the nucleon isospin doublet containing proton (P) and neutron ($n$) fields, the Goldstone bosons are parametrized by
 \begin{eqnarray}\label{U}
U(\pi) = u(\pi)^2 = \exp\left(  \frac{ i \boldpi \cdot \boldtau}{f_\pi}\right),
\end{eqnarray}
where $\boldpi$ is the pion triplet, $\boldtau$ the Pauli matrices, $f_\pi = 92.4$ MeV the pion decay constant, and $c_{1,5} \sim \mathcal O(1/\Lambda_\chi)$ are LECs associated to the nucleon sigma term and strong proton-neutron mass splitting. A hat denotes the traceless component of a chiral structure, e.g. $\hat \chi = \left(\chi  -\frac{1}{2} \textrm{Tr}(\chi)\right)$, and $\chi_\pm  =  u^{\dagger} \chi u^{\dagger} \pm u \chi^{\dagger} u\ .$

We can now read off the relevant interactions beginning with DM-pion interactions
\begin{eqnarray}\label{eq:pion}
\mathcal L_{\chi,q}^{\pi} = c^\pi_q \boldpi^2\,\bar \chi \chi\, , \qquad c^\pi_q = \frac{\mpi^2}{4}\left[ c_u (1-\varepsilon) + c_d (1+\varepsilon ) \right] \equiv \frac{\mpi^2}{2} \bar c_{q^{\mathrm{(is)}}} \, ,
\end{eqnarray}
where $\varepsilon = (m_d - m_u)/(m_d + m_u) =0.37\pm0.03$ \cite{Agashe:2014kda}, and we defined the effective isoscalar DM coupling $\bar c_{q^{\mathrm{(is)}}}$. 
Similarly we can read off the tree-level DM-nucleon interactions \cite{Crivellin:2013ipa}
\begin{eqnarray}\label{eq:nucleon}
\mathcal L_{\chi, N} &=& c^{N,\mathrm{(is)}}_q\bar N N\,\bar \chi \chi + c^{N,\mathrm{(iv)}}_q\bar N \tau^3 N\,\bar \chi \chi \, , \nonumber\\
c^{N,\mathrm{(is)}}_q&=&  -4 m_\pi^2 c_1 \bar c_{q^{\mathrm{(is)}}}\, , \nonumber\\
c^{N,\mathrm{(iv)}}_q&=& B (m_d - m_u) c_5  \left[c_u (1-\frac{1}{\varepsilon}) + c_d (1+\frac{1}{\varepsilon}) \right]\equiv B(m_d - m_u) c_5  \bar c_{q^{\mathrm{(iv)}}} , 
\end{eqnarray}
where $c^{N,\mathrm{(is)}}_q$ and $c^{N,\mathrm{(iv)}}_q$ are, respectively, the coupling strengths of the isoscalar and isovector DM-nucleon interactions, and we defined the effective isovector DM coupling $\bar c_{q^{\mathrm{(iv)}}}$. 

\subsubsection{Power counting}
\label{sec:power_counting}
\begin{figure}[t]
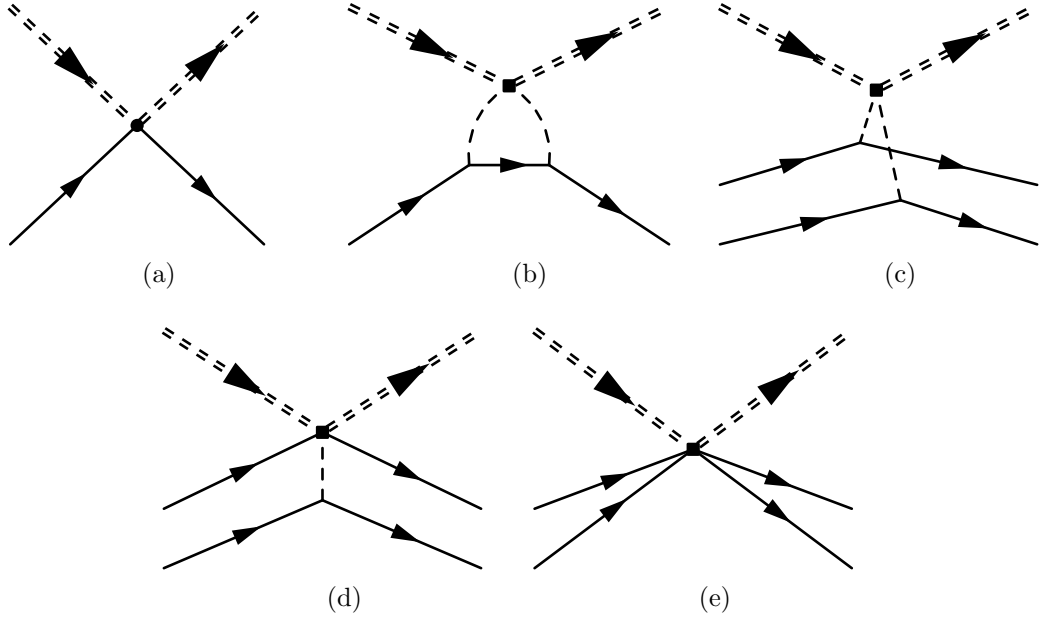

	\centering
	\begin{subfigure}[b]{0.3\textwidth}
		\input{feynmf/DM_2N_scalar.tex}
		\caption{\label{fig:diags:contact}}
	\end{subfigure}
	\begin{subfigure}[b]{0.3\textwidth}
		\input{feynmf/DM_2N_loop.tex}
			\caption{\label{fig:diag:loop}}
	\end{subfigure}
	\begin{subfigure}[b]{0.3\textwidth}
		\input{feynmf/DM_4N_2Pi.tex}
			\caption{\label{fig:diag:2nuc2pi}}
	\end{subfigure}
	\\ $\,$ \\
	\begin{subfigure}[b]{0.3\textwidth}
		\input{feynmf/DM_4N_1Pi.tex}
			\caption{\label{fig:diag:2nuc1pi}}
	\end{subfigure}
	\begin{subfigure}[b]{0.3\textwidth}
		\input{feynmf/DM_4N_0Pi.tex}
			\caption{\label{fig:diag:2nuc0pi}}
	\end{subfigure}
	\caption{\label{fig:isoscalar_diag}Diagrams contribution to DM-nucleus scattering.  Solid lines correspond to nucleons, single dashed lines to pions and double-dashed lines to DM.  Our analysis focuses on the diagrams in the first row, while those in the second row only appear at higher order.}
\end{figure}

The DM-nucleon interactions contribute to DM-nucleus scattering via Fig.~\ref{fig:diags:contact}, while the DM-pion interactions contribute via one-nucleon and two-nucleon interactions via Fig.~\ref{fig:diag:loop} and \ref{fig:diag:2nuc2pi}, respectively. A power-counting scheme is necessary to determine the relative order of these and other contributions. We count powers of the generic momentum P, where P is determined from the nuclear binding momentum which for typical nuclei is of the order $p\sim m_\pi $ as usual in $\chi$PT. In addition, we have the momentum transfer, $q$, between DM and the nucleus, which for scattering off light nuclei is expected to be somewhat smaller, but for simplicity we treat $q\sim p$. 

Weinberg showed \cite{Weinberg:1990rz} that the usual $\chi$PT power counting needs to be adapted for $A\geq 2$ intermediate states  that contain only propagating nucleons. A diagram can then be separated into two parts which do not contain such states (the irreducible part) and a part which does (the reducible part). Inside an irreducible subloop, the contour integration over the time component of the loop momentum can always be done in such a way that the nucleon pole is avoided and the nucleon energy is of order $\sim p$ as in standard $\chi$PT. An irreducible diagram can then be counted via the rules: $p^4/(4\pi)^2$ for each loop, $1/p$ for each nucleon propagator, $1/p^2$ for a pion propagator, and the product of the LECs associated to the relevant interactions. In irreducible diagrams, however, the nucleon poles cannot be avoided and the nucleon energy becomes $\sim p^2/m_N$ instead of $\sim p$. For such reducible diagrams we use the modified rules: $p^5/((4\pi)^2 m_N)$ for each loop, $m_N/p^2$ for each nucleon propagator, $1/p^2$ for a pion propagator, and the product of the LECs associated to the relevant interactions. Typically $p/m_N$ is counted as $p^2/\Lambda_\chi^2 \sim Q^2$, indicating a suppression of two orders in the chiral counting \cite{Epelbaum:2014efa}.

We can apply these rules to determine the hierarchy of the diagrams in Fig.~\ref{fig:isoscalar_diag}. We use $\bar c_{q^{\mathrm{(is)}}}$ and $\bar c_{q^{\mathrm{(iv)}}}$ defined in Eqs.~\eqref{eq:pion} and \eqref{eq:nucleon} instead of $c_u$ and $c_d$, as the former have a simpler power counting. As we do not know the (relative) sizes of $\bar c_{q^{\mathrm{(is)}}}$, $\bar c_{q^{\mathrm{(iv)}}}$, $c_s$, and $c_G$, we have to determine the power counting for each DM interaction at the quark-gluon level separately. For each DM interaction we denote the dominant contribution by $(0)$, the NLO by $(1)$, and so on. 

We begin with the isoscalar DM interactions. Diagram \ref{fig:diags:contact} is counted as as $c^{N,(is)}_q \sim m_\pi^2 c_1 \bar c_{q^{\mathrm{(is)}}} \sim \bar c_{q^{\mathrm{(is)}}} (m_\pi^2/\Lambda_\chi) $, where we took into account an overall normalization common to all diagrams.  This contribution forms the LO structure at order $(0)$.

Diagram \ref{fig:diag:loop} contains an additional irreducible loop, one nucleon propagator, two pion propagators, and two strong pion-nucleon vertices such that the relative scaling becomes
\begin{equation}
\mathcal A_b \sim c^\pi_q \left(\frac{p^4}{(4\pi)^2}\right)\left(\frac{1}{p}\right)\left(\frac{1}{p^2}\right)^2 \left(\frac{g_A \, p}{f_\pi}\right)^2 \sim c^\pi_q \frac{g_A^2\, p}{(4\pi f_\pi)^2} \sim  \bar c_{q^{\mathrm{(is)}}}  \frac{m_\pi^2}{\Lambda_\chi}\times \frac{p}{\Lambda_\chi}\, ,
\end{equation}
where we identified $4\pi f_\pi \sim \Lambda_\chi$ and counted $g_A \sim 1$. We obtain the familiar result \cite{Prezeau:2003sv,Cirigliano:2012pq}  for the isoscalar DM-quark interaction that the pion loop is suppressed by $p/\Lambda_\chi \sim Q$ and is therefore one order down in the chiral power counting. This contribution therefore appears at order $(1)$.  

We now turn to Diagram \ref{fig:diag:2nuc2pi} which, compared to Diagram \ref{fig:diags:contact}, has an additional reducible loop, one nucleon propagator, two pion propagators, and two strong pion-nucleon vertices. The power-counting predicts
\begin{equation}
\mathcal A_c \sim c^\pi_q \left(\frac{p^5}{(4\pi)^2 m_N} \right)\left(\frac{m_N}{p^2}\right)\left(\frac{1}{p^2}\right)^2 \left(\frac{g_A \, p}{f_\pi}\right)^2  \sim  \bar c_{q^{\mathrm{(is)}}} \frac{m_\pi^2}{\Lambda_\chi}\times \frac{p}{\Lambda_\chi}\, ,
\end{equation}
such that this diagram appears at order $(1)$ as well. Higher-order corrections are discussed in the next subsection.

We can perform a similar counting for isovector DM interactions. Diagram \ref{fig:diags:contact} is counted as as $c^{N,(iv)}_q \sim B(m_d -m_u) c_5 \bar c_{q^{\mathrm{(iv)}}} \sim \bar c_{q^{\mathrm{(iv)}}} (\varepsilon m_\pi^2/\Lambda_\chi) $, which provides the leading (0) contribution. It is tempting to argue that isovector DM interactions lead to small one-nucleon currents with respect to isoscalar DM because of the additional factor $\varepsilon$ appearing in the scaling of the diagram. However, such a comparison depends also on the ratio of $\bar c_{q^{\mathrm{(iv)}}}/\bar c_{q^{\mathrm{(is)}}}$ which depends on unknown parameters in the underlying DM model and can therefore be enhanced (see e.g. Ref.~\cite{Crivellin:2015bva}). We can therefore not compare contributions from different DM operators in a model-independent way.

 Since isospin-violating DM interactions do not lead to the DM-pion interactions in Eq.~\eqref{eq:pion},  Diagrams \ref{fig:diag:loop} and  \ref{fig:diag:2nuc2pi} do not appear at order (1), but only at higher orders in the power counting. For instance, operators like $\bar \chi \chi\,\pi_3^2$ appear after an additional insertion of the quark mass difference and are therefore suppressed by two powers in the chiral counting. Contributions from the diagrams with topology analogous to Figs. \ref{fig:diag:loop} and \ref{fig:diag:2nuc2pi} therefore only appear at order (3).

\subsubsection{LO and NLO currents for isoscalar and isovector DM-quark interactions}

The calculation of the first two diagrams in Fig.~\ref{fig:diag:loop} is straightforward. We consider a nucleon of incoming (outgoing) momentum P ($p'$) and define $\vec q = \vec p- \vec p^{\,\prime}$. 
Up to order (1), the nucleon-DM current can be written as
\begin{eqnarray}
J_{q^{(\mathrm{is} + \mathrm{iv})}} (\vec q) &=&  \left[\left(-4 m_\pi^2 c_1- \frac{9 g_A^2 \pi m_\pi^3}{4 (4\pi f_\pi)^2}\right) - \frac{9 g_A^2 \pi m_\pi^3}{4 (4\pi f_\pi)^2}F\left(\frac{|\vec q|}{2m_\pi}\right)\right] \,  \bar c_{q^{\mathrm{(is)}}}\nonumber \\
&& + B (m_d - m_u) c_5\,  \bar c_{q^{\mathrm{(iv)}}}\,\tau^3_i\, ,
\end{eqnarray}
where $\tau^3_i$ acts on the isospin of the interaction nucleon and 
we defined a function of the momentum transfer
\begin{equation}
F(x) = \frac{-x + (1+2x^2)\arctan x}{3x }\,,
\end{equation}
which for small momentum transfer becomes $F(x)  \simeq \frac{5}{9}x^2 + \dots$. As noted in Ref.~\cite{Crivellin:2013ipa}, the momentum-independent part of the isoscalar current can be identified as the first two terms in the chiral expansion of the nucleon sigma term $ \sigma_{\pi N} = \bar m (d m_N/d \bar m)$ where $\bar m = (m_u + m_d)/2$ is the average light quark mass. Similarly, the isovector current can be identified as the strong proton-neutron mass splitting $(m_n - m_p)^{\mathrm{strong}}\equiv \delta m_N =-4 B c_5 (m_d -m_u)$. We can therefore resum higher chiral orders \cite{Crivellin:2013ipa} by writing
\begin{eqnarray}\label{NLO1body}
J_{q^{(\mathrm{is} + \mathrm{iv})}} (\vec q) &=&  \left[\sigma_{\pi N} - \frac{9 g_A^2 \pi m_\pi^3}{4 (4\pi f_\pi)^2}F\left(\frac{|\vec q|}{2m_\pi}\right)\right]   \,\bar c_{q^{\mathrm{(is)}}} -\frac{\delta m_N}{4}\bar c_{q^{\mathrm{(iv)}}}\,\tau^3_i\, .
\end{eqnarray}
A big advantage of the resummation is that the nucleon sigma term and mass splitting have been precisely determined
\begin{equation}
\sigma_{\pi N} = (59.1 \pm 3.5)\,\mathrm{MeV}\ ,\qquad \delta m_N = ( 2.32\pm0.17)\, \mathrm{MeV}\, ,
\end{equation}
from, respectively, a Roy-Steiner analysis of pion-nucleon scattering \cite{Hoferichter:2015dsa} and lattice-QCD calculations \cite{Brantley:2016our}. Recent lattice-QCD calculations \cite{Durr:2015dna} find a smaller number for the nucleon sigma term than the Roy-Steiner analysis. The chiral prediction for the slope of the scalar form factor, the coefficient of $\vec q^{\,2}$, is  smaller by roughly $40\%$ than a determination based on a dispersive analysis \cite{Hoferichter:2012wf}, indicating that higher-order effects can be relevant for the momentum-dependent interactions. 

We now turn to the two-nucleon current depicted in Fig.~\ref{fig:diag:2nuc2pi} which becomes\
 \begin{equation}\label{NLO2body}
J_{q^{(\mathrm{is})},2b} (\vec q)=-m_\pi^2 \left(\frac{g_A}{2f_\pi}\right)^2 \frac{(\vec \sigma_1 \cdot \vec q_1)(\vec \sigma_2 \cdot \vec q_2)}{(\vec q_1^{\,2}+m_\pi^2)(\vec q_2^{\,2}+m_\pi^2)} \tau_1 \cdot \tau_2\,\bar c_{q^{\mathrm{(is)}}}\ ,
\end{equation}
where $\vec  q_i = \vec p^{\,\prime}_i - \vec p_i$ is the difference between the outgoing and incoming momentum of nucleon $i$, and $\sigma_i$ ($\tau_i$) the spin (isospin) of nucleon $i$.

At the next order in the chiral expansion, i.e. order $(2)$, there appear several new contributions. For instance, the isoscalar and isovector one-nucleon currents can be dressed by pion loops. However, all such contributions can be absorbed in the resummation of $\sigma_{\pi N}$, $\delta m_N$, and the radius corrections \cite{Martinprivate}. That is, there appear no genuinely new topologies. At order $(3)$, several new contributions appear. Particularly interesting for the isoscalar DM interactions are new two-nucleon contributions that arise, for example, from two-derivative DM-pion-pion interactions (e.g. LEC $l_4$ in Ref.~\cite{Gasser:1983yg}), DM-nucleon-pion interactions (e.g. LEC $d_{16}$ in Ref.~\cite{Fettes:2000gb}), and DM-nucleon-nucleon contact interactions. The relevant topologies of the latter two contributions are shown in Fig.~\ref{fig:diag:2nuc1pi} and \ref{fig:diag:2nuc0pi}. 
For the isovector DM interactions, the first two-nucleon corrections appear as well as discussed at the end of Sect.~\ref{sec:power_counting}.
The corrections at this order have not been fully calculated and will not be considered here.

\subsection{Currents for DM-strange and DM-gluon interactions}
As the interactions between DM and strange quarks and gluons in Eq.~\eqref{Eq1} are invariant under $SU(2)$ chiral-symmetry transformations, the hierarchy of hadronic interactions will be different compared to those arising from the couplings to light quarks.  In particular, the interaction between DM and pions\footnote{In $SU(3)$ $\chi$PT the strange interactions would lead to DM-kaon and DM-eta vertices \cite{Cirigliano:2012pq}. However, including dynamical strange mesons is not consistent with the $\chi$EFT expansion of nuclear forces that we apply here.} only arises at higher order and the main contributions arise from DM-nucleon interactions. These have been studied in many papers, see e.g. Refs \cite{Hill:2014yxa,Bishara:2016hek} and references therein, and here we summarize the results. For the interactions to strange quarks, the relevant matrix element is the strange sigma term $\sigma_s = m_s(d m_N/d m_s) = 40\pm 10$ MeV from an average of lattice-QCD calculations \cite{Junnarkar:2013ac}. For the momentum-dependent term, which cannot be obtained from $SU(2)$ $\chi$PT but instead requires $SU(3)$ $\chi$PT, we follow Ref.~\cite{Hoferichter:2016nvd} and use a dispersive extraction \cite{Hoferichter:2012wf}. This gives for the one-nucleon current    
\begin{eqnarray}\label{Strange1body}     
J_s (\vec q) &=&\left( \sigma_{s}  - \dot \sigma_s\,\vec q^{\, 2}   \right) c_s\, ,
\end{eqnarray}
where $\dot \sigma_s = 0.3 \pm 0.2\,\mathrm{GeV}^{-1}$ which, by comparison to the numerical value of the radius correction in Eq.~\eqref{NLO1body},  we label as an NLO correction.  Isovector currents and two-nucleon currents only appear at N${}^3$LO and are neglected. 

For the gluonic interaction in Eq.~\eqref{Eq1}, the trace anomaly gives \cite{Shifman:1978zn}
\begin{eqnarray}\label{Glue1body}
J_G (\vec q) &=&   c^{N,(\mathrm{is})}_G  =- c_G\frac{8 \pi}{9} \left[ m_N - \sigma_{\pi N} - \sigma_s \right] \, ,
\end{eqnarray}
where the $\sigma_{\pi N}$ and $\sigma_s$ are formally order $(2)$ corrections which we absorb into $m_N^G \equiv m_N - \sigma_{\pi N} - \sigma_s$.  

As for the interactions with strange quarks, momentum-dependent, isovector, and two-nucleon contributions only appear at order $(3)$. 
However, it was noticed in Ref.~\cite{Hoferichter:2016nvd} that two-nucleon currents provide numerically significant contribution to the nuclear structure functions of ${}^{132}$Xe. These two-nucleon corrections arise from the DM-pion interactions  
\begin{eqnarray}
\mathcal L_{\chi,G}^{\pi} = c_G\frac{8 \pi}{9}\bar \chi \chi\, \left[(\partial_\mu \boldpi)^2 - \frac{3}{2} m_\pi^2 \boldpi^2 \right]\, ,
\end{eqnarray}
where the LECs are fixed by a comparison of the QCD and $\chi$PT energy-momentum tensor \cite{Bishara:2016hek}. The power-counting rules outlined in Sect.~\ref{PC} indicate that  the resulting two-nucleon terms are indeed suppressed by three orders in the chiral expansion compared to the one-body contributions in Eq.~\eqref{Glue1body}. Nevertheless, we investigate the resulting two-nucleon contributions to estimate the size of the missing order $(3)$ terms and to study the numerical enhancement found in Ref.~\cite{Hoferichter:2016nvd}. We obtain for the two-nucleon current
 \begin{equation}\label{Glue2body}
J_{G, 2b} (\vec q)=- c_G \frac{8 \pi}{9}\left(\frac{g_A}{2f_\pi}\right)^2 \frac{(\vec \sigma_1 \cdot \vec q_1)(\vec \sigma_2 \cdot \vec q_2)}{(\vec q_1^{\,2}+m_\pi^2)(\vec q_2^{\,2}+m_\pi^2)} \tau_1 \cdot \tau_2\, \left(2\vec q_1\cdot \vec q_2 - 3 m_\pi^2\right)\ ,
\end{equation}
where it should be stressed that there appear additional contributions at this order. In fact, in the limit of $|\vec q_{1,2}| \gg m_\pi$ the two-nucleon current approaches a DM-nucleon-nucleon contact interaction which appears at the same order in the chiral counting and is, in general, necessary to absorb the divergence and associated cut-off dependence of Eq.~\eqref{Glue2body}. This implies that the calculation of the contribution from Eq.~\eqref{Glue2body} only provides a rough estimate of the size of higher-order corrections.

\section{Generation of nuclear wave functions and scattering matrix elements}\label{sec:waves}

\subsection{Matrix elements}
We consider scattering processes of the type 
$ \chi( \vec p_\chi ) + T( \vec p_T ) \rightarrow \chi( \vec p_\chi' ) + T( \vec p_T' )$, 
where $T$ denotes the target nucleus of mass $m_T$ consisting of $A$ nucleons.  We express the spin-independent cross sections in the non-relativistic limit as \cite{DelNobile:2013sia}
\begin{equation}
	\frac{\diff{\sigma}}{\diff{\vec q^2}}
	=
	\frac{1}{64 \pi}
	\frac{
		\abs{ \mathcal{M}_A  (\vec q^2 )}^2
	}{
		m_T^2 m_\chi^2 v_\chi^2
	}
	\ ,
\end{equation}
where $\vec q = \vec p_\chi - \vec p_\chi'$ is the momentum  transfer from DM to the target nucleus, $v_\chi$ the DM velocity, and $\mathcal{M}_A$ the spin-independent scattering matrix element.  We compute the scattering matrix depending on the DM current $\hat J$ and the wave function of the target nucleus $\ket{\Psi_T, j m_j}$ -- with distinct total spin $j$ and spin polarization $m_j$
\begin{equation}
	\label{eq:current_to_scattering_matrix}
	\abs{ \mathcal{M}_A ( \vec q ^2) }^2 = \frac{(2 m_T )^2( 2 m_\chi )^2}{2j +1}\sum_{m_j, m_j'= -j}^j
	\abs{
		\Braket{ \Psi_T, j m_j' | \ \hat{ J }(\vec q ^2) \ | \Psi_T, j m_j }
	}^2 \ .
\end{equation}
The mass factors arise from the normalization of the relativistic particle states.  In the following, we use that the internal properties of the wave functions are independent of the polarization $m_j$. The spin averaging guarantees that the matrix element is spherically symmetric such that we are free to choose the direction of the momentum transfer $\vec q$.  For convenience, we pick $\vec q = q \uvec e_z$.  As a consequence, $m_j$  is conserved and the scattering process is independent of the value of $m_j$. Hence, the sum cancels against the spin averaging.  We denote $\ket{\Psi_T} = \ket{\Psi_T, j m_j }$ from now on.  The properties of the applied wave functions are described in Sect.~\ref{subsec:waves}.

To obtain the differential recoil rates measured in experiments, we have to convolve the cross section with the DM velocity distribution in the Earth frame
\begin{equation}
	\frac{\diff{R}}{{\diff {\vec q^2}}}
	=
	\frac{1 }{m_T}
	\frac{ \rho_\chi }{m_\chi}
	\ \int \limits_{v_\chi^{(\mathrm{min})}}^{v_\chi^{(\mathrm{esc})}}
	\diff {^3 \vec v_\chi} \abs{ \vec v_\chi } f( \abs{ \vec v_\chi} ) \frac{\diff{\sigma}}{\diff{\vec q^2}}(\vec v_\chi)
	\ ,
\end{equation}
where $\rho_\chi$ is the local DM density,  $v_\chi^{(\mathrm{esc})} \approx 550\, \mbox{km}/\mbox{s}$ \cite{Smith:2006ym} is the maximal (escape) velocity which can be inferred from galaxy velocity distributions.  The momentum transfer $\vec q$, in the non-relativistic limit, is constrained through energy conservation by a product of the minimal DM velocity $v_\chi^{(\mathrm{min})}$ and the reduced mass of the colliding system, $\mu_T = m_T m_\chi /(m_T + m_\chi)$, times two.  Hence, the upper bound for the momentum transfer is given by
\begin{equation}
	\abs{\vec q}
	\lesssim
	2 \mu_T v_\chi^{(esc)}
	\lesssim
	A \times 2.5 \MeV
	\ ,
\end{equation}
for DM masses $\geq$ 1 GeV and for the light nuclei we consider here. For large nuclei the momentum transfer can become of the order of the pion mass and $\vec q^{\,2}$ corrections become more important.
Since one of our main goals is to estimate the accuracy of DM-nucleus cross sections, we present cross sections for larger $\vec q^{\,2}$ then relevant for scattering off light nuclei assuming the expected DM velocity distributions, in order to test whether the accuracy of our results depends on the momentum transfer. 

The scattering matrix element can be expressed in terms of a sum over different nuclear response functions $\mathcal{F}_{i, a}^{(\nu)}$
\begin{equation}\label{eq:defStructure}
	\abs{ \mathcal{M}_A (\vec q^2 ) }^2
	=
	(2 m_T )^2( 2 m_\chi )^2 \ \sigma_N^{(\mathrm{is})} \frac{\pi A^2 }{\mu_N^2} \abs{ \sum_{i,a,\nu} \alpha_i  \mathcal{F}_{i, a}^{(\nu)}(\vec q^{\,2}) }^2\ ,
\end{equation}
where we factored out $A^2$ and the LO isoscalar cross section of a DM-nucleon scattering process at zero momentum transfer
\begin{equation}\label{eq:defCross1N}
	\sigma_N^{(\mathrm{is})}
	=
	\frac{\mu_N^2}{\pi } \abs{ \sigma_{\pi N} \ c_{\chi} }^2
	\ .
\end{equation}
We have set here $\bar c_{q^{\mathrm{(is)}}}  = c_\chi$ and $\bar c_{q^{\mathrm{(iv)}}} = c_s = c_G=0$. This arbitrary choice is just for normalization purposes. $\alpha_i$ is a dimensionless coefficient which depends on the DM couplings. As mentioned before, we can expand $\mathcal{F}^{(\nu)}_{i,\,a}(\vec q^{\,2})$ order by order in the chiral expansion, but, since we cannot make model-independent statements about the (relative) sizes of the DM couplings $\bar c_{q^{\mathrm{(is)}}},\bar c_{q^{\mathrm{(iv)}}}, c_s, c_G$, we have a separate chiral power counting for each of the fundamental interactions.  If it is necessary to 	  consider several DM interactions at the same time then the power counting must be aligned depending on the relative sizes of the couplings under consideration. We discuss this in Sect.~\ref{sec:cancelation-regime} for the case of nonzero $\bar c_{q^{\mathrm{(is)}}}$ and $c_G$. Finally, we note that we recover the global counting of Ref.~\cite{Hoferichter:2016nvd} once we assume the various $c_i$ to be of the same size.

We will present our results in terms of response functions $\mathcal{F}^{(\nu)}_{i,\,a}(\vec q^{\,2})$ which are classified according to
\begin{itemize}
	\item $i= \{q^{(\mathrm{is})},\, q^{(\mathrm{iv})},\, s,\, G\}$ indicating, respectively, dependence on the various DM couplings we consider $\{\bar c_{q^{\mathrm{(is)}}},\, \bar c_{q^{\mathrm{(iv)}}},\, c_s,\, c_G \}$,
	\item at which chiral order $\nu = 0,\ 1,\dots$ in the chiral expansion of the currents they appear. The dominant current for each DM interaction, i.e. $\{\bar c_{q^{\mathrm{(is)}}},\, \bar c_{q^{\mathrm{(iv)}}},\, 		c_s,\, c_G \}$, starts at order $0$.
	\item We divide higher-order contributions into two-nucleon terms ($a=2b$) and one-nucleon radius ($\vec q^{\,2}$) corrections ($a= r$). 
\end{itemize}
We can now expand the structure functions appearing in the cross section as
\begin{align}
	\frac{\diff{\sigma}}{\diff{\vec q^2}}
	=
	\frac{ \sigma_N^{(\mathrm{is})} A^2}{4 \mu_N^2 v_\chi^2 }
	\bigg|
		&\quad\alpha_{q^{(\mathrm{is})}}  \left(
			\mathcal{F}^{\left(0\right)}_{q^{(\mathrm{is})}}{\left(\vec q^2\right)} + 
			\mathcal{F}^{\left(1\right)}_{q^{(\mathrm{is})},\ 2b}\left(\vec q^2\right)+
			\mathcal{F}^{\left(1\right)}_{q^{(\mathrm{is})},\ r}\left(\vec q^2\right) + 
			\dots 
		\right)\nonumber \\ & +
		\alpha_{q^{(\mathrm{iv})}} \left( 
			\mathcal{F}^{\left(0\right)}_{q^{(\mathrm{iv})}}{\left(\vec q^2\right)}  + 
			\dots 
		\right)+
		\alpha_{s} \left( 
			\mathcal{F}^{\left(0\right)}_{s}{\left(\vec q^2\right)} + 
			\mathcal{F}^{\left(1\right)}_{s,\, r}{\left(\vec q^2\right)}   + 
			\dots 
		\right)\nonumber \\ & \label{eq:def-response_functions} +
		\alpha_{G} \left( 
			\mathcal{F}^{\left(0\right)}_{G}{\left(\vec q^2\right)} + 
			\mathcal{F}^{\left(3\right)}_{G,\ 2b}\left(\vec q^2\right) + 
			\dots 
		\right)
		\bigg|^2\, .
		\end{align} 
		
Here we expanded the currents up to order (1) and since there appear no new corrections at order (2), the dots indicate missing order (3) corrections. The only order (3) correction we explicitly consider is the two-nucleon contribution arising from the DM-gluon interactions (see Eq.~\eqref{Glue2body}). We keep this term as a diagnostic tool to study the contributions of missing higher-order corrections. The dimensionless couplings $\alpha_i$ are given by
\begin{equation}\label{eq:alpha}
	\alpha_{q^{(\mathrm{is})}} = \frac{\bar c_{q^{\mathrm{(is)}}}}{c_\chi}, \qquad 
	\alpha_{q^{(\mathrm{iv})}} = \left(-\frac{\delta m_N}{4 \sigma_{\pi N}}\right) \frac{\bar c_{q^{\mathrm{(iv)}}}}{c_\chi}, \qquad 
	\alpha_s =  \left(\frac{\sigma_{s}}{ \sigma_{\pi N}}\right)\frac{\bar c_s}{c_\chi}, \qquad 	
	\alpha_G= \left(- \frac{8 \pi}{9} \frac{m^G_N}{\sigma_{\pi N}}\right) \frac{c_G}{c_\chi}\ .
\end{equation}
The factor of $c_\chi^{-1}$ is an artifact of the normalization in Eqs.~\eqref{eq:defStructure} and \eqref{eq:defCross1N}.
With these definition we obtain
\begin{eqnarray}\label{normalization}
	\mathcal{F}^{\left(0\right)}_{q^{(\mathrm{is})}}(\vec q^2)
	&=&
	\mathcal{F}^{\left(0\right)}_{s}(\vec q^2)
	\,\,=\,\,
	\mathcal{F}^{\left(0\right)}_{G}(\vec q^2)\, ,
\nonumber\\
	\mathcal{F}^{\left(0\right)}_{q^{(\mathrm{is})}}( 0) &=& \mathcal{F}^{\left(0\right)}_{s}(0)
	\,\,\,=\,\,\,
	\mathcal{F}^{\left(0\right)}_{G}(0) \,\,\,=\,\,\,  1\, .
\end{eqnarray}
The radius corrections $\mathcal{F}^{\left(1\right)}_{q^{(\mathrm{is})},\ r}\left(\vec q^2\right)$ and $\mathcal{F}^{\left(1\right)}_{s,\, r}{\left(\vec q^2\right)}$ involve the same nuclear matrix elements as the order (0) one-nucleon contributions apart from an additional overall dependence on $\vec q^2$. They therefore do not require additional nuclear calculations.
The independent structure functions are therefore $\mathcal{F}^{\left(0\right)}_{q^{(\mathrm{is})}}(\vec q^2)$, $\mathcal{F}^{\left(0\right)}_{q^{(\mathrm{iv})}}(\vec q^2)$, $\mathcal{F}^{\left(1\right)}_{q^{(\mathrm{is})},\ 2b}\left(\vec q^2\right)$, and $\mathcal{F}^{\left(3\right)}_{G,\ 2b}\left(\vec q^2\right)$ in agreement with the findings of Ref.~\cite{Hoferichter:2016nvd}. 

Although the ratio of LECs appearing in the definitions of $\alpha_i$ ranges from small values $-\delta m_N/(4 \sigma_{\pi N}) \simeq -0.01$ to large values $(-8\pi m^G_N)/(9\sigma_{\pi N}) \simeq -40$, this does not reflect the relative importance of the various terms. The $\alpha_i$ depend on the definition of the DM couplings $c_{u,d,s,G}$ in Eq.~\eqref{Eq1}. For instance, Eq.~\eqref{Eq1} assumes that the scalar couplings between light quarks and DM is proportional to the light-quark mass. While this is a reasonable assumption, it can be easily different in specific UV-complete models. We can only make model-independent statements about the relative sizes of different contributions proportional to the same $\alpha_i$.

\subsection{Nuclear wave functions}\label{subsec:waves}

Our calculations are based on a momentum-space evaluation of matrix elements that involve 
the DM interactions as introduced in the previous section and nuclear wave functions. Below, we require 
wave functions for $^{2}$H and $^{3}$He (and ${}^3$H) which we obtain from solutions of the non-relativistic Schr\"odinger
equation in momentum space. 

In the case of the deuteron, we directly solve 
\begin{equation}
 | \psi_{d} \rangle = \frac{1}{E_{d}-T} \ V_{12} \     | \psi_{d} \rangle \ .
\end{equation}
Here, $E_{d}$ is the deuteron binding energy, $T$ the two-nucleon (NN) kinetic energy and $V_{12}$ the NN 
interaction. The deuteron wave function is expanded in momentum eigenstates $| \, p \, \alpha \rangle$ where 
$\alpha$ corresponds to the partial waves contributing to the deuteron bound state: the orbital angular momentum 
$l_{12}=0,2$, NN spin $s_{12}=1$ and the total angular momentum $j_{12}=1$. 

We consider both, modern phenomenological NN interactions and interactions based on chiral effective theory 
to obtain the wave functions. The two standard choices for phenomenological interactions are AV18 \cite{Wiringa:1995co}
and CD-Bonn \cite{Machleidt:2001ib}. Although both interactions describe the available NN data below the pion 
production threshold essentially perfectly, their properties are quite different. The CD-Bonn interaction is much 
more non-local than AV18. For the deuteron, the non-observable kinetic energies, potential energies and 
D-state probabilities  differs visibly for both models as can be seen from Table~\ref{tab:deuteron}. 
Therefore, a comparison of results for these models is a good indication of possible model dependences. 
For the radius and the quadrupole moment, there are slight deviations from the experimental values. These 
can be traced back to neglected contributions, e.g. meson-exchange currents and relativistic corrections, 
and uncertainties in the correction of the measured charge radii for the finite radii of the proton and neutron.

\begin{table}
\scriptsize
\begin{tabular}{l|lll|ll|lll}
NN interaction           & $E_{d}$ & $\langle T \rangle $ &$\langle V \rangle$& $r_{d}$ &   $Q_{d}$ & $ P_{D} $ & $A_{S}$  & $\eta$ \\
\hline \hline
AV18                         & -2.225    &  19.81                      &  -22.04                  &   1.967  &      0.270 &        5.76   & 0.884       &  0.0252  \\ 
CD-Bonn                   & -2.223    &  15.60                      & -17.83                   &   1.966  &      0.270 &        4.85   & 0.884       &  0.0258 \\
\hline
LO\,\,\,\,\,\, $(Q^0)$ $\Lambda_{1}$  & -1.989   & 14.26   &  -16.25   &   1.997   &   0.245   &    3.27     &     0.825    &    0.0219    \\
LO\,\,\,\,\,\, $(Q^0)$   $\Lambda_{2}$  & -2.023   & 13.29   &  -15.32   &   1.990   &   0.230   &    2.54     &     0.833    &    0.0212    \\
LO\,\,\,\,\,\, $(Q^0)$   $\Lambda_{3}$  & -2.083   & 12.47   &  -14.55   &   1.979   &   0.215   &    1.97     &     0.849    &    0.0205    \\
LO\,\,\,\,\,\, $(Q^0)$   $\Lambda_{4}$  & -2.167   & 11.76   &  -13.92   &   1.965   &   0.199   &    1.53     &     0.870    &    0.0198    \\
LO\,\,\,\,\,\, $(Q^0)$   $\Lambda_{5}$  & -2.272   & 11.15   &  -13.42   &   1.950   &   0.183   &    1.18     &     0.897    &    0.0192    \\
\hline 
NLO\,\, ($Q^{2}$)   $\Lambda_{1}$  & -2.191   & 15.40   &  -17.59   &   1.970   &   0.275   &    5.23     &     0.875    &    0.0256    \\
NLO\,\, ($Q^{2}$)   $\Lambda_{2}$  & -2.199   & 14.25   &  -16.45   &   1.968   &   0.273   &    4.73     &     0.877    &    0.0256    \\
NLO\,\, ($Q^{2}$)   $\Lambda_{3}$  & -2.206   & 13.49   &  -15.69   &   1.967   &   0.271   &    4.24     &     0.879    &    0.0257    \\
NLO\,\, ($Q^{2}$)   $\Lambda_{4}$  & -2.211   & 12.92   &  -15.14   &   1.965   &   0.269   &    3.77     &     0.881    &    0.0258    \\
NLO\,\, ($Q^{2}$)   $\Lambda_{5}$  & -2.213   & 12.48   &  -14.69   &   1.965   &   0.267   &    3.35     &     0.881    &    0.0259    \\
\hline 
N${}^2$LO ($Q^{3}$)   $\Lambda_{1}$  & -2.228   & 14.94   &  -17.16   &   1.967   &   0.271   &    4.87     &     0.886    &    0.0254    \\
N${}^2$LO ($Q^{3}$)   $\Lambda_{2}$  & -2.231   & 13.85   &  -16.08   &   1.966   &   0.270   &    4.50     &     0.886    &    0.0256    \\
N${}^2$LO ($Q^{3}$)   $\Lambda_{3}$  & -2.235   & 13.17   &  -15.40   &   1.964   &   0.270   &    4.12     &     0.888    &    0.0258    \\
N${}^2$LO ($Q^{3}$)   $\Lambda_{4}$  & -2.237   & 12.69   &  -14.93   &   1.964   &   0.269   &    3.75     &     0.888    &    0.0260    \\
N${}^2$LO ($Q^{3}$)   $\Lambda_{5}$  & -2.235   & 12.32   &  -14.55   &   1.963   &   0.269   &    3.40     &     0.887    &    0.0263    \\
\hline 
N${}^3$LO ($Q^{4}$)   $\Lambda_{1}$  & -2.223   & 23.33   &  -25.55   &   1.970   &   0.268   &    3.78     &     0.884    &    0.0255    \\
N${}^3$LO ($Q^{4}$)   $\Lambda_{2}$  & -2.223   & 21.58   &  -23.80   &   1.972   &   0.271   &    4.19     &     0.884    &    0.0255    \\
N${}^3$LO ($Q^{4}$)   $\Lambda_{3}$  & -2.223   & 19.63   &  -21.85   &   1.975   &   0.275   &    4.77     &     0.885    &    0.0256    \\
N${}^3$LO ($Q^{4}$)   $\Lambda_{4}$  & -2.223   & 17.71   &  -19.94   &   1.979   &   0.279   &    5.21     &     0.885    &    0.0256    \\
N${}^3$LO ($Q^{4}$)   $\Lambda_{5}$  & -2.223   & 16.13   &  -18.35   &   1.982   &   0.283   &    5.58     &     0.885    &    0.0256    \\
\hline 
N${}^4$LO ($Q^{5}$)   $\Lambda_{1}$  & -2.223   & 20.64   &  -22.86   &   1.970   &   0.271   &    4.28     &     0.884    &    0.0256    \\
N${}^4$LO ($Q^{5}$)   $\Lambda_{2}$  & -2.223   & 18.89   &  -21.12   &   1.972   &   0.271   &    4.29     &     0.884    &    0.0256    \\
N${}^4$LO ($Q^{5}$)   $\Lambda_{3}$  & -2.223   & 17.48   &  -19.70   &   1.974   &   0.272   &    4.40     &     0.884    &    0.0255    \\
N${}^4$LO ($Q^{5}$)   $\Lambda_{4}$  & -2.223   & 16.29   &  -18.51   &   1.978   &   0.276   &    4.74     &     0.885    &    0.0256    \\
N${}^4$LO ($Q^{5}$)   $\Lambda_{5}$  & -2.223   & 15.27   &  -17.50   &   1.981   &   0.280   &    5.12     &     0.885    &    0.0256    \\
\hline \hline
Expt.                                  &   -2.225  \cite{VanDerLeun:1982gh} & --- & --- &  1.975 \cite{Huber:1998fp} &    0.286(2) \cite{Bishop:1979zz}  & --- &  0.878(4) \cite{Borbely:1985tur} &  0.0256(4) \cite{Rodning:1990hm}
\end{tabular}
\caption{\label{tab:deuteron}Properties of the deuteron wave functions used. The deuteron binding energy $E_{d}$, the 
expectation value of the kinetic energy and of the potential are given in MeV. We also give the point proton rms radius 
$r_{d}$ in fm, the quadrupole moment $Q_{d}$ in fm$^{2}$ and the D-state probability $P_{D}$ in \% together with the 
asymptotic normalization $A_{S}$ in fm$^{-1}$ and $\eta = \frac{A_{D}}{A_{S}}$.} 

\end{table}

For the chiral interactions, we use the ones of Ref.~\cite{Epelbaum:2014sza}. Following the approach of Weinberg \cite{Weinberg:1990rz,Weinberg:1992jd},
the non-perturbative character of the nuclear interactions is taken into account by expanding a nuclear potential
perturbatively in terms of the pions mass $m_{\pi}$ and the nucleon 3-momentum  $q$ over the chiral symmetry breaking or breakdown 
scale  $\Lambda_{b}$.  This potential is then used in a Schr\"odinger equation to obtain observables of nuclear systems. 

We note that there has been some discussion how such a non-perturbative expansion can be made consistently in leading 
and higher orders \cite{Birse:2006jn,Nogga:2005hb,PavonValderrama:2006gp,PavonValderrama:2005uj,Epelbaum:2006pt,Valderrama:2011hw,Valderrama:2011hz,Epelbaum:2009hn}. 
The approach used applies all orders of the interaction non-perturbatively. This has been shown to result in a high-quality 
description of NN observables provided the interactions are regulated using finite values of a cut-off. For the interactions used here  
the regularization is defined in configuration space and parameterized by a short-distance scale $R$. Explicit calculations have shown 
that $R$ needs to be larger than $0.8$~fm. The best description of the NN data is obtained for $R=0.9$~fm. Based on the 
description of NN scattering data, the breakdown scale was estimated in \cite{Epelbaum:2014sza} and turns out to be 
dependent on the cut-off chosen. We have summarized the available cut-off values in Table~\ref{tab:cut-offs} where we also introduce 
the short hand notation $\Lambda_{i}$ for the various cut-off values. To finalize our discussion of the NN interactions, we note that 
AV18 is accompanied by an electromagnetic (EM) interaction which we also take into account. Additionally, we have added the 
same EM interaction to the proton-proton and neutron-neutron interactions in case  of CD-Bonn and the chiral 
interactions. In the latter cases, we did not add the  EM part to the $np$ interactions, since this would lead to a visible deviation of the deuteron binding 
energy to experiment. However, at least for CD-Bonn, the EM contribution is consistent since the interaction has been 
fitted to NN data taking the EM contributions into account. We stress that the EM contribution is small and of minor importance. 

\begin{table}
\centering
\begin{tabular}{c | c c c c c}
	cut-off    & $\Lambda_1$ & $\Lambda_2$ & $\Lambda_3$ & $\Lambda_4$ & $\Lambda_5$ \\ \hline
	R   & 0.8 fm        & 0.9 fm      & 1.0 fm       & 1.1fm      & 1.2 fm     \\
	$\Lambda_b$ & 600 MeV & 600 MeV         & 600  MeV       & 500  MeV       & 400 MeV
\end{tabular}
\caption{\label{tab:cut-offs}cut-off values used to compute wave functions and the uncertainty estimates.  Values are taken from Ref.~\cite{Binder:2016bi}. }
\end{table}

In Table~\ref{tab:deuteron}, we also summarize the properties of the deuteron wave function for the chiral interactions at different 
orders of the chiral expansion $Q^{i}=(q,m_{\pi} / \Lambda_{b})^{i}$, see Sect.~\ref{uncertainty} for more details. It is  seen that non-observable quantities like 
the D-wave probability, the kinetic and the potential energy strongly  depend on the cut-off value chosen. Radii, binding energies, quadrupole moments, 
and asymptotic normalizations show much less variation and, in higher orders, reproduce the experimental values similarly to the 
phenomenological interactions. We stress, however, that this does not necessarily imply that we can expect results for DM scattering that are independent 
of the wave function. 

In order to obtain the wave functions for $^{3}$He (and ${}^3$H), we rewrite the Schr\"odinger equation into Faddeev equations
\begin{equation}
| \psi_{12} \rangle = G_{0} \ t_{12} \ P \ | \psi_{12} \rangle + G_{0} \ \left( \ 1+ t_{12} G_{0} \ \right) \  V_{123}^{(3)} \ (1+P)  | \psi_{12} \rangle \, ,
\end{equation}
where we have introduced the Faddeev components $| \psi_{12} \rangle$. Due the antisymmetry of the $^{3}$He wave function, 
all Faddeev components can be related to $| \psi_{12} \rangle$ using permutation operators.  
The combination of transpositions $P_{ij}$  that enters the Faddeev equations is labeled by $P=P_{12}P_{23} + P_{13}P_{23}$.  $G_{0}$ 
denotes the 3N free propagator. NN interactions enter via the t-matrices $t_{12}$ which are obtained by solving 
a Lippmann-Schwinger equation for the NN system embedded into a 3N system. Finally, for some of the phenomenological calculations, we  take 3NFs into 
account. These are first separated in three parts again related by permutation operators $V_{123}=V_{123}^{(1)}+V_{123}^{(2)}+V_{123}^{(3)}$. 
Only one of these parts is required for the Faddeev equation. The equations are solved using Jacobi relative momenta. 
Therefore the basis depends on two momenta, the relative momentum in the subsystem $(12)$ of the first two nucleons $p_{12}$ and the relative 
momentum $p_{3}$ of the third particle with respect to the other two. The angular dependence is expanded in corresponding orbital angular 
momenta $l_{12}$ and $l_{3}$. As in the NN system above, the orbital angular momentum $l_{12}$ is coupled with the spin of the NN system $s_{12}$ 
to the total two-body angular momentum $j_{12}$. The orbital angular momentum  $l_{3}$ is coupled with the spin of the third nucleon to a spectator
angular momentum $I_{3}$. $j_{12}$ and $I_{3}$ are finally combined to the total angular momentum $j_{3}=1/2$. 

From the isospin $t_{12}$ of the (12) 
subsystem and the isospin of the third nucleon, we build the total isospin $\tau_{3}$ of $^{3}$He. Since we take the charge dependence 
of the nuclear force and electromagnetic forces into account, both isospins $\tau_3=1/2$ and $3/2$ contribute although the 3/2 component 
is small \cite{Nogga:2003iy}. For this work, we do not take the proton/neutron mass difference into account but assume an average 
nucleon mass of $m_{N}=938.9182$~MeV. The number of partial waves states is constrained by the maximal NN angular momentum $j^{max}_{12}$. 
We chose  $j^{max}_{12}=6$ for the phenomenological interactions and $j^{max}_{12}=7$ for the chiral ones.  With these constraints on the 
partial waves and using approximately 60 momentum grids points for $p_{12}$ and $p_{3}$, we were able to obtain binding energies to an accuracy of 
1~keV. More details on the calculations can be found in Ref.~\cite{Nuclearandhypernuc:2001wd}. 

\begin{table}
\scriptsize
\begin{tabular}{l|lll|lll|lll|l}
NN interaction           & $E(^{3}{\rm He})$ & $\langle T \rangle $ &$\langle V \rangle$& $r_{p}$ & $r_{n}$ &  $r_{NN}$ &   $P_{S}$ & $ P_{P} $ & $ P_{D} $  & $ \langle \Psi  |   \Psi \rangle $ \\
\hline \hline
AV18                                  & -6.922  & 45.67   &  -52.59 &  1.871  &    1.677    &  3.127      &      91.47       &    0.065     &   8.465 &  0.998752 \\ 
AV18 + Urb IX                    &  -7.754  & 50.21   &  -57.97 &  1.770  &    1.601    &  2.966     &      90.63       &    0.131     &   9.241 &  0.998956 \\
CD-Bonn                            &  -7.264  & 36.77   &  -44.03 &  1.819   &   1.637    &  3.047     &      92.95       &    0.046     &   7.000 &  0.999505  \\
CD-Bonn + TM                   & -7.729   & 38.53  &  -46.26  &  1.767   &   1.598    &  2.964     &      92.41       &    0.090     &   7.498 &  0.999589  \\
\hline
LO\,\,\,\,\,\, $(Q^0)$ $\Lambda_{1}$  & -11.22   & 54.29   &  -65.51   & 1.327  &    1.258   &   2.259     &       95.31     &     0.031    &    4.659 & 0.999959  \\
LO\,\,\,\,\,\, $(Q^0)$  $\Lambda_{2}$  & -10.92   & 48.71   &  -59.63   & 1.367  &    1.289   &   2.324     &       96.65     &     0.015    &    3.333 & 0.999969  \\
LO\,\,\,\,\,\, $(Q^0)$  $\Lambda_{3}$  & -10.47   & 43.40   &  -53.87   & 1.424  &    1.334   &   2.416     &       97.58     &     0.008    &    2.410 & 0.999972  \\
LO\,\,\,\,\,\, $(Q^0)$  $\Lambda_{4}$  & -10.01   & 38.70   &  -48.70   & 1.489  &    1.388   &   2.522     &       98.23     &     0.004    &    1.763 & 0.999976  \\
LO\,\,\,\,\,\, $(Q^0)$  $\Lambda_{5}$  & -9.594   & 34.67   &  -44.27   & 1.558  &    1.444   &   2.634     &       98.69     &     0.002    &    1.304 & 0.999979  \\
\hline 
NLO\,\, ($Q^{2}$)   $\Lambda_{1}$  & -7.315   & 36.61   &  -43.92   & 1.805  &    1.626   &   3.025     &       92.19     &     0.057    &    7.755 & 0.999781  \\
NLO\,\, ($Q^{2}$)   $\Lambda_{2}$  & -7.481   & 33.96   &  -41.44   & 1.784  &    1.610   &   2.992     &       93.06     &     0.046    &    6.890 & 0.999881  \\
NLO\,\, ($Q^{2}$)   $\Lambda_{3}$  & -7.638   & 32.75   &  -40.38   & 1.764  &    1.595   &   2.960     &       93.94     &     0.037    &    6.025 & 0.999924  \\
NLO\,\, ($Q^{2}$)   $\Lambda_{4}$  & -7.804   & 32.21   &  -40.01   & 1.744  &    1.579   &   2.927     &       94.76     &     0.029    &    5.210 & 0.999946  \\
NLO\,\, ($Q^{2}$)   $\Lambda_{5}$  & -7.969   & 31.96   &  -39.93   & 1.723  &    1.563   &   2.895     &       95.51     &     0.022    &    4.466 & 0.999959  \\
\hline 
N${}^2$LO ($Q^{3}$)   $\Lambda_{1}$  & -7.305   & 35.68   &  -42.99   & 1.821  &    1.639   &   3.050     &       92.99     &     0.045    &    6.966 & 0.999771  \\
N${}^2$LO ($Q^{3}$)   $\Lambda_{2}$  & -7.409   & 32.88   &  -40.29   & 1.806  &    1.628   &   3.027     &       93.61     &     0.039    &    6.357 & 0.999877  \\
N${}^2$LO ($Q^{3}$)  $\Lambda_{3}$  & -7.540   & 31.63   &  -39.17   & 1.788  &    1.614   &   2.999     &       94.26     &     0.032    &    5.713 & 0.999925  \\
N${}^2$LO ($Q^{3}$)   $\Lambda_{4}$  & -7.680   & 31.13   &  -38.81   & 1.769  &    1.599   &   2.969     &       94.90     &     0.027    &    5.078 & 0.999947  \\
N${}^2$LO ($Q^{3}$)   $\Lambda_{5}$  & -7.824   & 30.96   &  -38.79   & 1.749  &    1.584   &   2.937     &       95.51     &     0.022    &    4.470 & 0.999960  \\
\hline 
N${}^3$LO ($Q^{4}$)   $\Lambda_{1}$  & -6.865   & 51.81   &  -58.67   & 1.884  &    1.691   &   3.148     &       94.79     &     0.027    &    5.189 & 0.998593  \\
N${}^3$LO ($Q^{4}$)   $\Lambda_{2}$  & -6.871   & 50.74   &  -57.61   & 1.890  &    1.697   &   3.159     &       94.07     &     0.032    &    5.897 & 0.998904  \\
N${}^3$LO ($Q^{4}$)   $\Lambda_{3}$  & -6.832   & 46.94   &  -53.77   & 1.901  &    1.707   &   3.180     &       93.12     &     0.041    &    6.842 & 0.999219  \\
N${}^3$LO ($Q^{4}$)   $\Lambda_{4}$  & -6.812   & 42.31   &  -49.13   & 1.910  &    1.716   &   3.196     &       92.43     &     0.048    &    7.524 & 0.999481  \\
N${}^3$LO ($Q^{4}$)   $\Lambda_{5}$  & -6.779   & 38.21   &  -44.99   & 1.921  &    1.725   &   3.215     &       91.89     &     0.053    &    8.052 & 0.999661  \\
\hline 
N${}^4$LO ($Q^{5}$)   $\Lambda_{1}$  & -6.790   & 52.31   &  -59.10   & 1.898  &    1.701   &   3.170     &       93.96     &     0.033    &    6.005 & 0.998833  \\
N${}^4$LO ($Q^{5}$)   $\Lambda_{2}$  & -6.893   & 47.33   &  -54.22   & 1.886  &    1.693   &   3.154     &       93.92     &     0.033    &    6.046 & 0.999185  \\
N${}^4$LO ($Q^{5}$)   $\Lambda_{3}$  & -6.931   & 43.30   &  -50.23   & 1.885  &    1.695   &   3.155     &       93.71     &     0.035    &    6.255 & 0.999424  \\
N${}^4$LO ($Q^{5}$)   $\Lambda_{4}$  & -6.919   & 39.84   &  -46.76   & 1.893  &    1.702   &   3.169     &       93.17     &     0.040    &    6.786 & 0.999597  \\
N${}^4$LO ($Q^{5}$)   $\Lambda_{5}$  & -6.872   & 36.75   &  -43.62   & 1.906  &    1.714   &   3.192     &       92.60     &     0.046    &    7.351 & 0.999719  \\
\hline \hline
Expt.                                  &  -7.718 \cite{Audi:2012dv,Wang:2012fh} & ---        & ---           & 1.776  & ---  & ----             & ---                &  ---            &   ---       &   ---     
\end{tabular}
\caption{\label{tab:3he}Properties of the $^3$He wave functions used. The $^{3}$He binding energy $E(^{3}{\rm He})$, the 
expectation value of the kinetic energy and of the potential are given in MeV. We also give the point proton and neutron rms radius 
and the rms distance of two nucleons, $r_{p}$,  $r_{n}$  and $r_{NN}$, respectively, in fm.  The S-, P-, and D-state probabilities $P_{S}$, $P_{P}$, and $P_{D}$ are given in \%. 
The deviations of the  norm of the wave functions from 1 are a measure of higher partial-wave contributions (see text). } 

\end{table}

The binding energy results together with some basic wave function properties are summarized in Table~\ref{tab:3he}. The first observation is that the binding 
energies are generally not in very good agreement with the experimental value. This can be attributed to contributions of three-nucleon forces (3NFs). For the 
phenomenological interactions, we have added 3NFs that have been tuned to describe the mirror nucleus $^{3}$H correctly \cite{Nogga:2002kw}. 
Thereby, AV18 was augmented by the Urbana~IX interaction (Urb~IX) \cite{Pudliner:1997fv} and CD-Bonn with the Tucson-Melbourne (TM) force \cite{Coon:2001it}. 
Although it is an important strength of chiral interactions that they are accompanied by 3NFs that can be derived within the same 
framework, we have not added an chiral 3NFs to the chiral interactions in this work since their adjustment is work in progress. We show later 
that our results here are not significantly affected by Urb~IX and TM. We therefore expect that adding the chiral 3NFs is not too important for the results in this work but we stress that this needs to be verified in a future calculation.

The table also gives results for  point proton and neutron radii together with the rms distance of two nucleons. The rather strong dependence 
on the interaction model is due to a strong correlation with the binding energy \cite{Friar:1987hr} and therefore driven by the long-distance 
components of the wave function. We compare the point proton radius to an experimental value that is based on the charge 
radius of $^{3}$He of $r_{ch}= 1.976(15)$~fm \cite{Ottermann:1985km} and corrected for the finite proton and neutron size along the lines 
explained in Ref.~\cite{Binder:2016bi}.

 We also give the probability to find total orbital angular momentum $L=0,1,2$ in our wave functions. To obtain 
these numbers, we performed a recoupling of angular momenta to an LS coupling scheme. Besides the dominant S-wave component, we 
find a sizable D-wave component. The P-wave component is quite small. 

The last column of the table gives results for the norm of the wave function. It can be shown using the antisymmetry 
of the wave function that the scalar product of the wave function $ |   \Psi \rangle = (1+P) |   \Psi_{12} \rangle$ is
\begin{equation}
\langle \Psi  |   \Psi \rangle  = 3 \langle \psi_{12}  |   \Psi \rangle \ . 
\end{equation}
Numerically, this relation is not strictly fulfilled since  higher partial waves are missing. We use the right hand side 
for the normalization of the wave function. This is generally leads to a larger accuracy since the 
partial wave convergence of the Faddeev component is faster then that of the wave function. The last column 
shows the left-hand side of the equation. The deviation from one is small, but still, some higher partial wave 
contributions are missing. Their contribution to matrix elements is generally suppressed when finite-range 
operators are considered as we do in this work. 

\subsection{Chiral expansion and uncertainty estimation}\label{uncertainty}

The calculation of the scattering amplitude, or equivalently the response functions, depends on both the nuclear wave function ($\Psi$) and the DM current ($\hat J$) that is sandwiched between the wave functions.  Both quantities can be expanded order by order in the chiral expansion. 
The LO contribution to the cross section arises from combining LO wave functions with the order (0) one-nucleon current.  For isoscalar DM-quark interactions the first corrections arise from applying NLO wave functions and from the order (1) two-nucleon and radius corrections to the currents.  For the DM-strange quark interactions the currents are only altered by the radius corrections, while for the remaining DM interactions only the wave function is affected. Analogously, even higher-order wave functions should be combined with higher-order corrections to the currents. However, whereas we have access to N${}^3$LO and N${}^4$LO wave functions, we do not control the currents at the same order. That is, a consistent calculation with these wave functions would also require the calculation of missing higher-order currents. This can be done with the methods of Refs.~\cite{Baroni:2015uza, Krebs:2016rqz}, but is beyond the scope of this work.

As discussed below, we find that the corrections associated to the inclusion of higher order-currents are smaller than the corrections associated with higher-order wave functions.  For this reason, we chose to present our uncertainty estimation based on the variation of different orders in the wave functions but fixed current orders: we sandwich the one- and two-nucleon currents derived in Sect.~\ref{PC} between wave functions calculated from LO up to N${}^4$LO nucleon-nucleon potentials.  We show that at least N${}^3$LO wave functions are required in order to get precise results for two-nucleon currents.

To assign an uncertainty to our calculation at each order in the chiral expansion of the wave function we follow the scheme of Ref.~\cite{Binder:2016bi} which we slightly modify in order to account for missing high-order effects of the currents.  The objective of this scheme is to make a prediction for the value of the observable $X$ with a well-defined uncertainty estimate. In this work, the observable $X$ corresponds to the magnitude squared of the response functions $X = \abs{\mathcal{F}}^2$, but the procedure can be used for any quantity.  As we are only able to approximate $X$ with wave functions at finite chiral order $\nu$, it is desirable to quantify the uncertainty $\delta^{(\nu)} X$ associated with the approximation.  The calculation of $X^{(\nu)}$ is uncertain up to missing corrections proportional to $Q^{\nu + 1}$
\begin{equation}
	X^{(\nu)} = X + \delta^{(\nu)} X = X + \sum _{i =1}^\infty Q^{\nu+i} \, \Delta X^{(\nu)}_i\ \,,
\end{equation}
where $X$ would be the result of a fictitious infinite-order calculation. $\Delta X^{(\nu)}_i$ denotes a set of coefficients parameterizing the uncertainty for a given order in $Q$ -- the dimensionless expansion parameter that is given by the generic momentum $p \sim m_\pi \sim q$ divided by the large breakdown scale. For the numerical error estimates we use $Q = \mathrm{max}(q/\Lambda_b,\, m_\pi/\Lambda_b)$ with $\Lambda_b$ a regulator scale which depends on the coordinate-space cut-off used in the evaluation of the nuclear wave functions (see Table \ref{tab:cut-offs}). That is, for small momentum transfers, $q$ is determined by $m_\pi/\Lambda_b$. 

For a well-behaved expansion, all the coefficients $\Delta X^{(\nu)}_i$ are expected to be of natural order $\Delta X^{(\nu)}_i \sim X$.  We approximate the error by assuming that one can find a maximal coefficient 
$ \Delta X^{(\nu)}_{\max} \geq | \Delta X^{(\nu)}_{i} | $, which allows us to resum the series 
\begin{align}
	\delta^{(\nu)} X 
	\leq
	Q^{\nu} \abs{ \Delta X^{(\nu)}_{\max} } \sum _{i =1}^\infty Q^{i}
	=
	\frac{Q^{\nu + 1 }}{1 - Q} \abs{ \Delta X^{(\nu)}_{\max} }
	\ .
\end{align}
This uncertainty  estimate is rather conservative as we have used not only the maximal term but also assumed a coherent summation of uncertainties.  To find the size of the coefficient, we analyze the difference of the same observable at two different chiral orders $\nu' > \nu$
\begin{align}
	X^{(\nu)} - X^{(\nu')}
	=
	Q^{\nu} \left( \Delta X^{(\nu)}_1 + \mathcal{ O }(Q) \right)
	\ .
\end{align}
We finally estimate this coefficient by further taking the maximal value over all computed differences at different chiral orders
\begin{align}
	\abs{ \Delta X^{(\nu)}_{\max} } \leq \abs{ \Delta X_{\max} }
	&= \max \limits_{\nu} \left( \abs{ \Delta X^{(\nu)}_{\max} } \right) \ ,
	&
	\Delta X^{(\nu)}_{\max}
	&\simeq
	\max_{\nu' > \nu} \left(
		\frac{ \abs{ X^{(\nu)} - X^{(\nu')} } }{Q^{\nu +1}}
	\right)
	\ .
\end{align}
The final uncertainty estimate is then given by
\begin{equation}
	\delta X^{(\nu)}
	\leq
	\frac{Q^{\nu + 1 }}{1 - Q} \abs{ \Delta X_{\max} }
	\ .
\end{equation}

While this method provides a conservative estimate of the uncertainty at a given chiral order of the wave functions, we stress again that the estimate does not fully capture the missing higher-order currents. In addition, the simple procedure outlined here does not provide a statistical interpretation of the theoretical uncertainty.

\section{Results and discussion}\label{sec:results}
\subsection{Convergence and uncertainty estimates}

We now turn to the results of our calculation. We mainly discuss the case of the isoscalar quark-DM interactions as they lead to the most interesting higher-order currents. The other cases are briefly discussed in Sect.~\ref{sec:others}. As discussed above, we calculate the structure functions that appear in Eq.~\eqref{eq:def-response_functions} for different chiral and phenomenological wave functions.  When using chiral wave functions we present results for each order in the chiral expansion and for different values of the cut-off used to regulate the bound-state equations.  The theoretical uncertainty of our result at each order in the chiral expansion is obtained by the method described in Sect.~\ref{uncertainty}. In this work we focus on the nuclear aspects of DM direct detection and we therefore do not show the uncertainty associated to hadronic quantities such as the uncertainty on $\sigma_{\pi N}$ and $\delta m_N$. We refer to, for example, Ref.~\cite{Cirigliano:2013zta} for discussions of the latter. We also do not include astrophysical uncertainties related to DM direct detection \cite{McCabe:2010zh}.

We begin with discussing the results for scattering off the deuteron. In the top-panel of Fig.~\ref{fig:err_deut} we show results for $|\mathcal{F}^{\left(0\right)}_{q^{(\mathrm{is})}}{\left(\vec q^2\right)}|^2$ for three different values of the momentum transfer $q=|\vec q|$. At zero momentum transfer the results are equal to unity for each applied wave function due to the normalization of the structure functions, see Eq.~\eqref{normalization}. For larger values of the momentum transfer the structure function decreases, which can be described by a Helm form factor as discussed in the next section.  We find essentially no dependence on the order of the chiral wave function used nor on the applied cut-off which is reflected by the very small uncertainty bands on the results.  

In the second row we show the effect of including the higher-order currents arising from radius and two-body corrections 
\begin{equation}
|\mathcal{F}^{\left(0+1\right)}_{q^{(\mathrm{is})}}{\left(\vec q^2\right)}|^2 \equiv |\mathcal{F}^{\left(0\right)}_{q^{(\mathrm{is})}}{\left(\vec q^2\right)} + \mathcal{F}^{\left(1\right)}_{q^{(\mathrm{is}),2b}}{\left(\vec q^2\right)}+ \mathcal{F}^{\left(1\right)}_{q^{(\mathrm{is}),r}}{\left(\vec q^2\right)}|^2\ .
\end{equation} 
Compared to the top panel, the results have shifted by only a small amount indicating that higher-order currents only moderately modify the cross section. The uncertainty of the total result, however, has increased significantly in particular for NLO chiral wave functions. As we discuss in more detail below, the increase in uncertainty is caused by a relatively large dependence of the two-body corrections on the applied wave function. Finally, we note that the relative effect of the higher-order currents seems to decrease somewhat for larger values of the momentum transfer. This happens because of mutual cancellations between the radius and two-nucleon corrections.

In the last two rows of Fig.~\ref{fig:err_deut} we study the effects of higher-order currents in more detail. In the third and fourth row we show, respectively, the relative contribution of the radius and two-body currents with respect to the full result shown in the second row. That is,
\begin{eqnarray}\label{Deltas}
\Delta^{(r)} &=& \frac{|\mathcal{F}^{\left(0+1\right)}_{q^{(\mathrm{is})}}{\left(\vec q^2\right)}|^2 - |\mathcal{F}^{\left(0\right)}_{q^{(\mathrm{is})}}{\left(\vec q^2\right)}+\mathcal{F}^{\left(1\right)}_{q^{(\mathrm{is}),2b}}{\left(\vec q^2\right)}|^2}{|\mathcal{F}^{\left(0+1\right)}_{q^{(\mathrm{is})}}{\left(\vec q^2\right)}|^2} \,, \nonumber \\ 
\Delta^{(2b)} &=& \frac{|\mathcal{F}^{\left(0+1\right)}_{q^{(\mathrm{is})}}{\left(\vec q^2\right)}|^2 - |\mathcal{F}^{\left(0\right)}_{q^{(\mathrm{is})}}{\left(\vec q^2\right)}+\mathcal{F}^{\left(1\right)}_{q^{(\mathrm{is}),r}}{\left(\vec q^2\right)}|^2}{|\mathcal{F}^{\left(0+1\right)}_{q^{(\mathrm{is})}}{\left(\vec q^2\right)}|^2} \, .
\end{eqnarray}
Just like the order (0) structure functions, the radius corrections do not depend on the order of the chiral wave function nor on the applied cut-off. This can be easily understood as the current has the same form as the order (0) one-body current apart from the dependence on $F(|\vec q|/2 m_\pi) \simeq (5/9) |\vec q^2|/(4 m_\pi^2) + \dots$ (see Eq.~\eqref{NLO1body}). The nuclear aspects of the calculation are therefore identical. The radius correction vanishes at zero momentum transfer and is only $-2\%$ for $q=100$ MeV. 

More interesting are the two-body corrections. In contrast to one-body currents we obtain a significant dependence on the applied wave function. If we apply NLO wave functions the effects of two-body currents at zero momentum transfer range from $\Delta^{(2b)} = (2\pm15)\%$ when using cut-off $\Lambda_1$ up to $2\pm3\,\%$ when using cut-off $\Lambda_2$, while the other cut-offs give values in between these extremes. This large uncertainty for all cut-offs indicates that the two-body corrections depend on aspects of the wave function not captured by the observables given in Table~\ref{tab:deuteron}. For instance, the NLO wave functions already give good agreement with experiments as far as the binding energy, radius, quadrupole moment, and the asymptotic $D/S$ ratio are concerned. Increasing the order of the chiral wave functions we observe an order-by-order convergence of the results. Using N${}^2$LO wave functions we still observe a significant cut-off dependence of the result ranging from $\Delta^{(2b)} = (3\pm4)\%$ for $\Lambda_1$ up to $\Delta^{(2b)} = (1.6\pm0.8)\%$ for $\Lambda_2$, but for even higher-order wave functions almost all cut-off dependence has disappeared and we obtain stable results. This can be seen in more detail from Table~\ref{tab:amp_res} given in Appendix \ref{appendix:2pi_res}. 

In line with previous observations \cite{Epelbaum:2014efa}, we find that using the cut-off $\Lambda_2$, gives rise to the fastest convergence pattern. Since all results for different $\Lambda_i$ are consistent once the most accurate wave functions are applied, in the following sections we will show results using $\Lambda_2$.

\MyIncFig{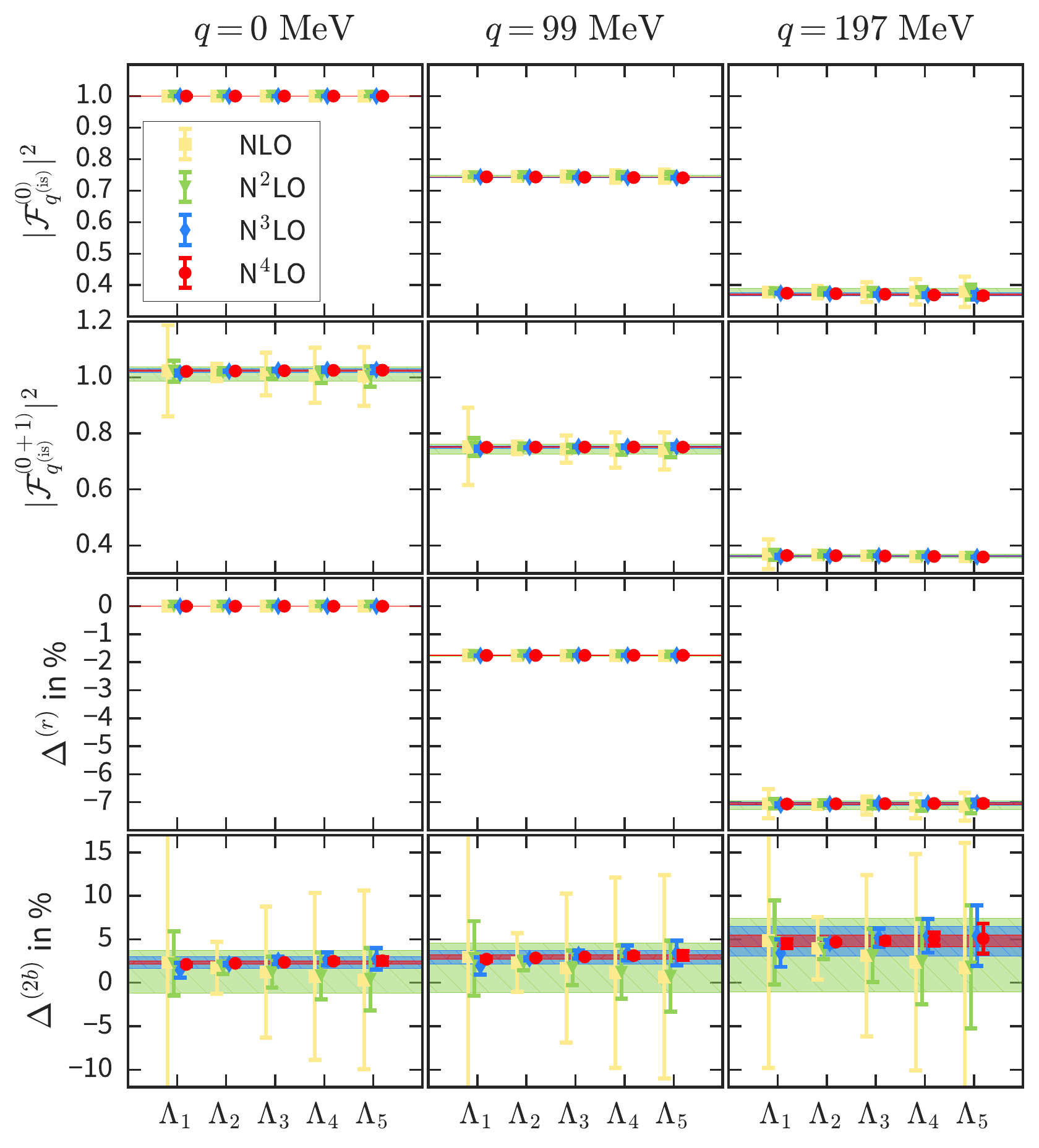}{width=\textwidth}{fig:err_deut}{The plots show different structure functions for isoscalar quark-DM interaction for the ${}^2$H system at three values of the momentum transfer $q$.  From top to bottom the panels indicate, respectively, the leading, order (0), one-nucleon isoscalar structure functions, the full order (0+1) isoscalar structure functions including two-body and radius corrections, the modification in percent due to the radius corrections, and the modification in percent due to two-body corrections. Results are shown for different chiral wave functions (NLO up to N${}^4$LO) and for the five different cut-offs given in Table~\ref{tab:cut-offs}. The colored bands, from outermost to innermost, correspond to averages over all cut-offs at N$^2$LO, N$^3$LO and N$^4$LO in the wave expansion. Results for LO wave functions are not shown because the uncertainty bands are too large.}

In Fig.~\ref{fig:err_he3} we show the analogous plots for scattering off ${}^3$He. We do not show the results for ${}^3$H as the main features are essentially the same (see Table \ref{tab:amp_res} for numerical results of the two-nucleon matrix elements.)
Regarding one-nucleon currents, the main difference with respect to the ${}^2$H case is that already the leading structure function has a significant uncertainty at larger momentum transfer even for N${}^2$LO wave functions. This uncertainty reflects the more complicated nature of three-nucleon wave functions.

The two-body contributions are smaller than in the deuteron case and significantly more uncertain. The relative two-body correction is roughly a factor $5$ smaller than for ${}^2$H. This is perhaps unexpected as the relative importance of two-body currents is naively expected to grow with $A$ compared to the one-body results. The smallness for ${}^3$He is probably related to the spin-isospin structure of ${}^3$He and was also observed in an analysis of pion-nucleus scattering lengths where very similar two-body currents appear \cite{Liebig:2010ki,Baru:2011bw}. We expect two-body currents to grow in denser nuclei and aim to investigate ${}^4$He in a forthcoming study. 

Independent of the size, it is interesting to study how accurate we can calculate the two-body corrections. Like for the ${}^2$H scenario we see that using NLO wave functions gives rise to significant uncertainties. The accuracy improves with N${}^2$LO wave functions, but unlike the ${}^2$H case, even with the fastest-converging cut-off, $\Lambda_2$, the uncertainty on the two-body correction is roughly $100\%$. At least N${}^3$LO wave functions are necessary to obtain results distinguishable from zero. 

\MyIncFig{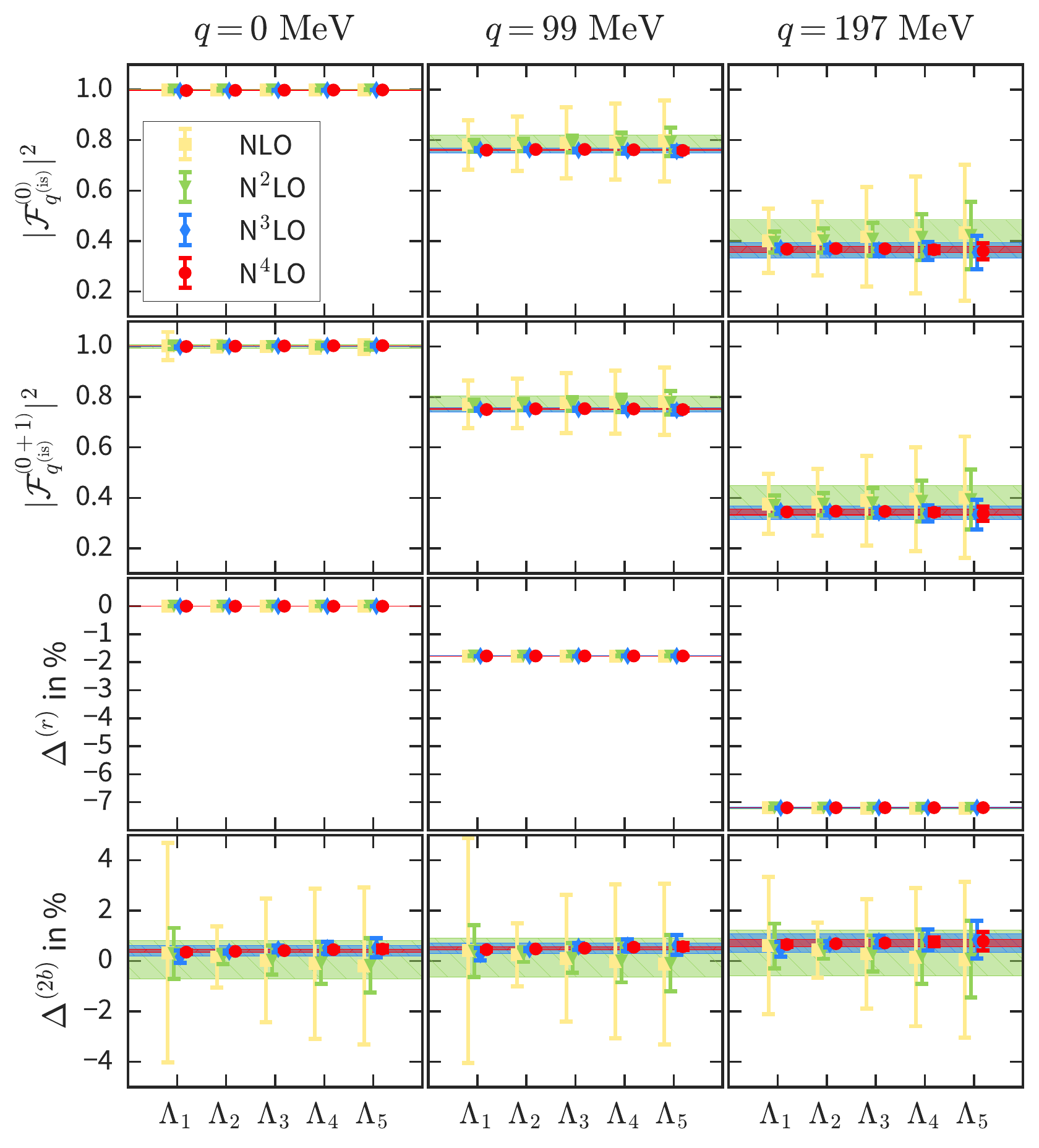}{width=\textwidth}{fig:err_he3}{The plots showsdifferent structure functions for isoscalar quark-DM interaction for the ${}^3$He system at three values of the momentum transfer $q$.  From top to bottom the panels indicate, respectively, the leading order (0)  one-body isoscalar structure functions, the full order $(0+1)$ isoscalar structure functions including two-body and radius corrections, the modification in percent due to the radius corrections, and the modification in percent due to two-body corrections. Results are shown for different chiral wave functions (NLO up to N${}^4$LO) and for the five different cut-offs given in Table~\ref{tab:cut-offs}. The colored bands, from outermost to innermost, correspond to averages over all cut-offs at N$^2$LO, N$^3$LO and N$^4$LO in the wave expansion.}

\subsection{Discussion}

\MyIncFig{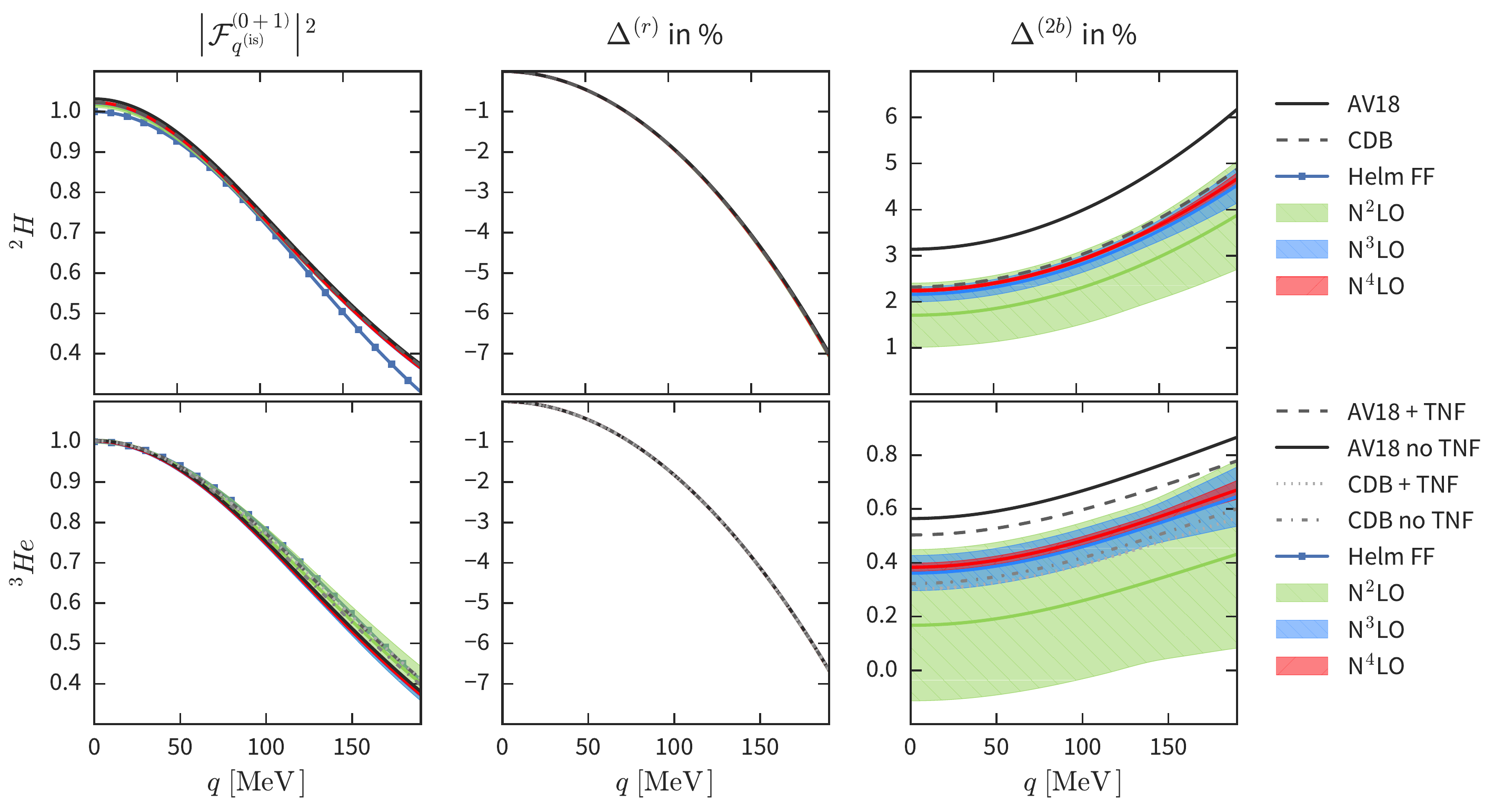}{width=\textwidth}{fig:con_quark}{The left column shows the full order (0+1) isoscalar-quark DM structure functions for ${}^2$H (top) and ${}^3$He (bottom).  The second and third column denote, respectively, the radius and two-nucleon corrections as defined in Eq.~\eqref{Deltas}. The uncertainty bands correspond to chiral wave functions at N$^2$LO, N$^3$LO and N$^4$LO (from outer- to innermost: green, blue, red). Results from using phenomenological wave functions and for the fitted Helm form factor (left panel only) are also shown.  The phenomenological wave functions for ${}^3$He are generated with and without three-nucleon forces (TNF).}

\begin{table}[t]
\centering
\begin{tabular}{c | cc}
	Target & $a$ (fm) & $s$ (fm) \\ \hline
	$^2$H  & 0.47 & 1.09  \\ 
	$^3$H  & 0.38 & 0.96  \\ 
	$^3$He & 0.39 & 0.98  \\ 
\end{tabular}
\caption{\label{tab:HFF}Helm form factor coefficients for the order (0) response functions for ${}^2$H and ${}^3$He.  The coefficients are fitted to $\mathcal{F}^{\left(0\right)}_{q^{(\mathrm{is})}}$ for N$^2$LO wave functions at cut-off $\Lambda_2$.}
\end{table}

We summarize our results in Fig.~\ref{fig:con_quark} where we show the full isoscalar structure functions, the radius corrections, and the two-body corrections obtained from  N$^2$LO-N$^4$LO wave functions using the fastest converging cut-off $\Lambda_2$. We also show results using phenomenological wave functions. As is clear from the first panel and the discussions above, the NLO corrections to the isoscalar structure functions for both ${}^2$H and ${}^3$He are small and the impulse approximation, i.e. neglecting all two-nucleon contributions, is excellent. The $\vec q^2$-dependence of the structure functions can be parametrized by a Helm form factor \cite{HelmFF,Lewin1996} (although this is usually done for heavier nuclei) which we fitted to our results for the leading one-nucleon response functions. We parametrize the form factor as
\begin{eqnarray}
	\mathcal{F}^{\left(\mathrm{H}\right)}( q )
	&=&
	3 \frac{j_1(q r_n)}{q r_n} \
	e^{- (q s )^2 / 2 } \ , \nonumber\\
	r_n^2 &=& c^2 + \frac{7}{3} \pi^2 a^2 - 5 s^2\ , \nonumber\\
	c &=& (1.23 A^{1/3} - 0.60) \fm\, ,
\end{eqnarray}
and provide fit values for $a$ and $s$ in Table~\ref{tab:HFF}. The fitted form factors describe the full results well over the considered range of momentum transfer. The result is also very stable with respect to different nucleon-nucleon potentials and, essentially, all wave functions give the same results. This is not surprising as the full structure function is dominated by the one-body scalar currents, which were found to be wave-function independent. The same holds true for the radius corrections shown in the second column of Fig.~\ref{fig:con_quark}. 

The effects of two-body currents are summarized in the third panel of Fig.~\ref{fig:con_quark}. Their magnitude is very modest in the light nuclei under consideration and in general cases it should be safe to neglect them. The fact that they only provide a few-percent correction is not in good agreement with the expected size based on chiral-EFT power counting, which predicts effects of $\mathcal O(p/\Lambda_\chi) \sim \mathcal O(m_\pi /\Lambda_\chi) \sim (10$-$30)\%$. 
Despite the small size of two-nucleon contributions, it is interesting to study the accuracy with which they can be determined as they can become more relevant in heavier nuclei \cite{Hoferichter:2016nvd} (as was found in the case of neutrino-nucleus scattering \cite{Lovato:2014eva}), in cases where one-body contributions are cancelled against other contributions \cite{Prezeau:2003sv, Cirigliano:2012pq}, or for non-scalar  DM interactions that lead to spin-dependent DM-nucleus scattering \cite{Klos:2013rwa}. Furthermore, the scalar two-nucleon currents at zero momentum transfer provide the dominant two-nucleon contribution to nuclear sigma terms \cite{Beane:2013kca}, which describe the dependence of the nuclear mass on the quark masses. 

Whereas the single-nucleon contributions are very stable, we find much a larger wave-function dependence for the two-nucleon contributions. Using N${}^2$LO chiral wave functions we observe a $50\%$ uncertainty on the ${}^2$H two-body matrix elements and $100\%$ on the ${}^3$He matrix elements. Both uncertainties decrease significantly once N${}^3$LO wave functions are applied. At this and higher chiral order the uncertainty estimates do not include the effects of potential higher-order currents. We briefly discuss these in the next section.

Results using high-order chiral wave functions are in good agreement with phenomenological wave functions. Where AV18 predicts somewhat larger two-body corrections, CD-Bonn is in very good agreement with the results for N${}^3$LO and N${}^4$LO wave functions. The main advantage of applying chiral wave functions is  that they allow for a systematic uncertainty estimate. Finally, our ${}^3$He chiral wave functions do not include three-nucleon forces which brings in a further uncertainty not quantified by the uncertainty bands. We expect however that three-nucleon forces do not affect the results significantly as they only mildly affect the results for phenomenological wave functions. Furthermore, the spread of phenomenological results with and without three-nucleon forces is of the same size as the estimated uncertainty of our N$^3$LO results.

\begin{figure}
\center
\includegraphics[width=.7\textwidth]{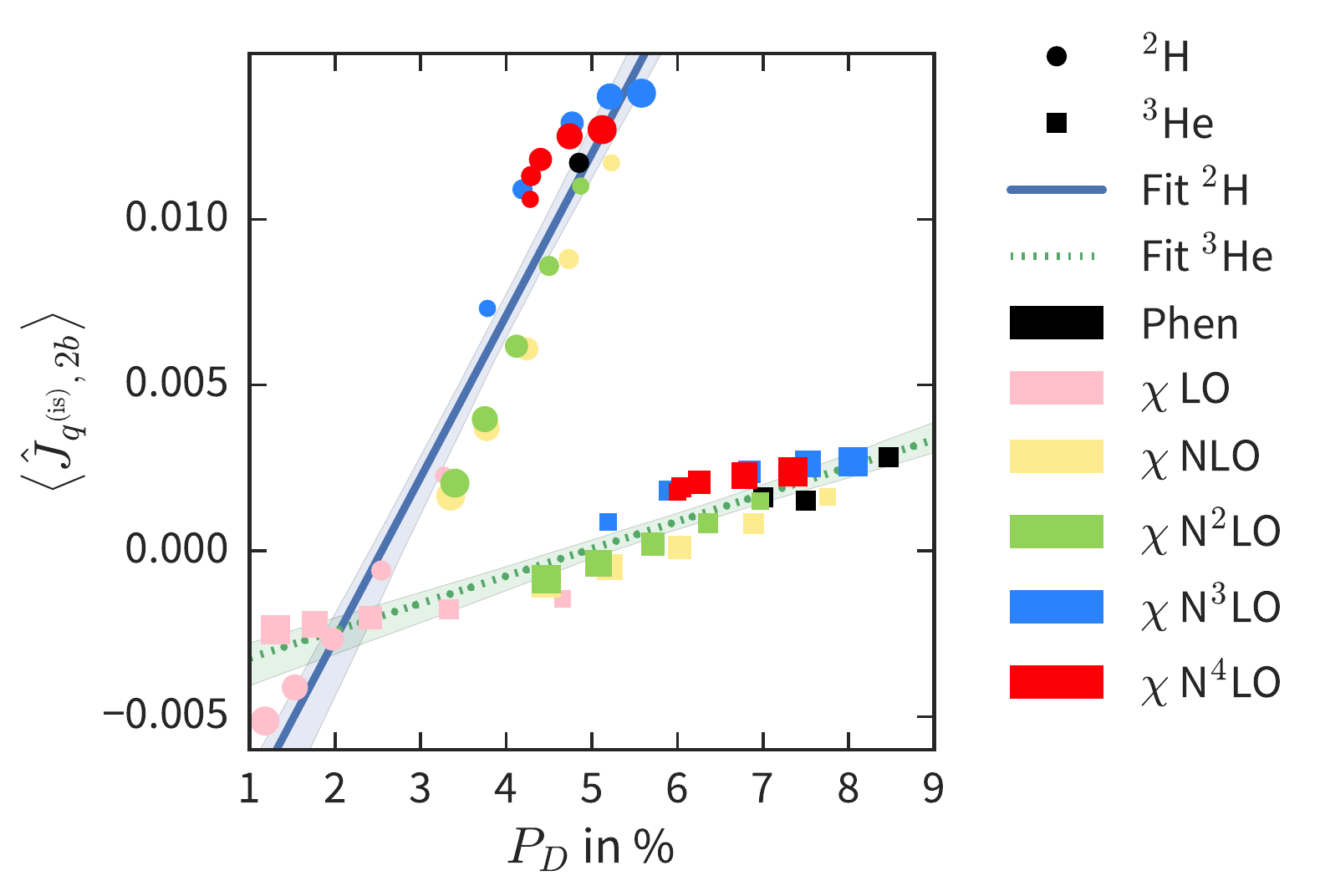}
\caption{Correlation between isoscalar two-nucleon matrix elements and the D-wave probability.  Circles and squares correspond to ${}^2$H and $^{3}$He, respectively.  The sizes of the circles and squares correspond to the sizes off the cut-off -- the smallest (largest) symbols represent $\Lambda_1$ ($\Lambda_5$).  Different colors correspond to different wave functions. A linear fit results in $\braket{J_{q^{(\mathrm{is})},2b}} = 4.9 \cdot 10^{-3} P_D -1.2\cdot 10^{-2}$  for ${}^2$H and $\braket{J_{q^{(\mathrm{is})},2b}} = 8.2 \cdot 10^{-4} P_D -4.1\cdot 10^{-3}$ for ${}^3$He.}
\label{fig:D_wave_correlation}
\end{figure}

It is interesting to study in more detail why the two-body current matrix elements suffer from large uncertainties even though the NLO and N${}^2$LO ${}^2$H and ${}^3$He wave functions describe several observable quantities quite well (see Tables~\ref{tab:deuteron} and \ref{tab:3he}). While studying the two-body current matrix elements we observed a significant correlation between the D-wave probability of the wave function and the two-nucleon matrix element (defined in Eq.~\eqref{def:op_expectation_value}). This correlation is shown in Fig.~\ref{fig:D_wave_correlation}. It turns out that the correlation follows an almost linear behavior with different slopes and interceptions for  ${}^2$H and ${}^3$He. The strong dependence of the two-body correction on the D-wave probability is also seen in the context of pion-deuteron scattering \cite{Liebig:2010ki,Baru:2011bw}.  The D-wave probabilities change significantly from N${}^2$LO to N${}^3$LO chiral wave functions, see Tables \ref{tab:deuteron} and \ref{tab:3he}, probably due to the appearance of new short-range  interactions in the nucleon-nucleon potential. The connection to the D-wave probability might also be interesting for lattice-QCD calculations of nuclear sigma terms at non-physical pion masses \cite{Beane:2013kca} as larger pion masses can lead to a suppressed nucleon-nucleon tensor force and thus a reduced D-wave probability.

The observation that two-body corrections suffer from large uncertainties when using N${}^2$LO chiral wave functions for light nuclei might indicate that the corresponding uncertainties for more complex heavier systems can be even larger, although further study is required to confirm this. This is potentially relevant for two-nucleon corrections in other contexts such as calculations of neutrino scattering off ${}^4$He and ${}^{12}$C nuclei \cite{Lovato:2014eva} using N${}^2$LO chiral wave functions obtained with Quantum Monte Carlo methods. In any case, it is worthwhile to note that wave functions with very similar properties regarding observables such as binding energies and electromagnetic moments, can still lead to different two-nucleon matrix elements.

\subsection{The remaining DM interactions}\label{sec:others}
Having discussed in detail the isoscalar quark DM-interactions, we now briefly turn to the remaining terms in Eq.~\eqref{Eq1}. As is clear from the parametrization of the differential cross section in Eqs.~\eqref{eq:def-response_functions}-\eqref{normalization}, the order (0) one-nucleon contributions from strange quark-DM and gluon-DM interactions depend on the same structure function as the isoscalar quark-DM interactions. The order (0) results can therefore be directly read off from the top panels of Figs.~\ref{fig:err_deut} and \ref{fig:err_he3}. The differential cross sections only differ in their overal size by the dependence on different DM couplings and strong LECs as indicated in Eq.~\eqref{eq:alpha}.
Similarly, the order (1) radius corrections for the strange quark-DM interactions can be read off from the third panel of Figs.~\ref{fig:err_deut} and \ref{fig:err_he3} after appropriate rescaling of the $\vec q^2$-dependence of the strange sigma term with respect to the light-quark sigma term (see Eqs.~\eqref{NLO1body} and \eqref{Strange1body}). These radius corrections are not present for isovector quark-DM or gluon-DM interactions until order (3) currents. 
Furthermore, none of the remaining DM interactions, $\bar c_{q^{\mathrm{(iv)}}}$, $c_s$, $c_G$, lead to two-nucleon corrections until order (3). We discuss one such correction below.

The only nontrivial difference with respect to the isoscalar structure functions then arises from the isospin structure $\tau_i^3$ that appears in Eq.~\eqref{NLO1body} for the leading one-nucleon current arising from $\bar c_{q^{\mathrm{(iv)}}}$. At zero momentum transfer the isoscalar operator counts the number of nucleon giving rise to the factor $A^2$ in Eq.~\eqref{eq:defStructure}. Similarly, at $\vec q^2=0$, the isovector interactions count the difference between the number of protons and neutrons. This implies that these contributions vanish for the DM-${}^2$H scattering but contribute to DM-${}^3$He (and ${}^3$H) scattering. 

 To investigate the typical size of isospin-violating corrections, we choose for the DM couplings in Eq.~\eqref{Eq1} the following values $c_u = c_d = c_\chi$ and $c_s=c_G=0$. This choice implies  $\alpha_{q^{\mathrm{(is)}}} =1$ and $\alpha_{q^{\mathrm{(iv)}}} =-\delta m_N/(2 \sigma_{\pi N}) \simeq -0.02$. Because of the smallness of $\alpha_{q^{\mathrm{(iv)}}}/\alpha_{q^{\mathrm{(is)}}}$, in this scenario the dominant contribution to the scattering process arises from the one-nucleon isoscalar current\footnote{Clearly other choices for $c_u$ and $c_d$ can lead to drastically different relative contributions. For instance, the choice $c_u = -c_d = c_\chi$ would enhance the ratio of $\alpha_{q^{\mathrm{(iv)}}}/\alpha_{q^{\mathrm{(is)}}}$ by a factor $\varepsilon^{-2} \simeq O(10)$.}
 Subleading corrections then arise from the isoscalar radius and two-nucleon corrections and the isovector contributions. The differential cross section is proportional to
 \begin{eqnarray}
 |\mathcal{F} (\vec q^2)|^2 &=& \bigg |\alpha_{q^{(\mathrm{is})}}  \left[ \mathcal{F}^{\left(0\right)}_{q^{(\mathrm{is})}}{\left(\vec q^2\right)} + 
			\mathcal{F}^{\left(1\right)}_{q^{(\mathrm{is})},\ 2b}\left(\vec q^2\right)+\mathcal{F}^{\left(1\right)}_{q^{(\mathrm{is})},\ r}\left(\vec q^2\right)\right] + \alpha_{q^{(\mathrm{iv})}} \mathcal{F}^{\left(0\right)}_{q^{(\mathrm{iv})}}{\left(\vec q^2\right)} \bigg |^2\, .
\end{eqnarray}

In Fig.~\ref{fig:log_plots_light} we plot the various contributions to $|\mathcal F(\vec q^2)|^2$ for scattering off ${}^2$H (left panel) and ${}^3$He (right panel) applying N${}^2$LO chiral wave functions using the fastest converging cut-off. As mentioned, the main contribution arises from the isoscalar contribution which is depicted by $\mathcal F^{(0)}_{q^{\mathrm{(is)}}}$. For scattering of ${}^2$H, there are no isospin-violating corrections and the first correction arises from the interference between the isoscalar one- and two-nucleon contributions depicted by $\mathcal F^{(0)}_{q^{\mathrm{(is)}}}-\mathcal F^{(1)}_{q^{\mathrm{(is)}},\,2b}$.  As discussed in more detail above, the two-nucleon corrections suffer from significant uncertainties illustrated by the red band. The final corrections come from the order (1) one-nucleon radius corrections and are smaller then the two-nucleon corrections over the considered range of momentum exchange. We do not show contributions from the square of two-nucleon or radius corrections (or their interference) as these are smaller by orders of magnitude. 

\MyIncFig{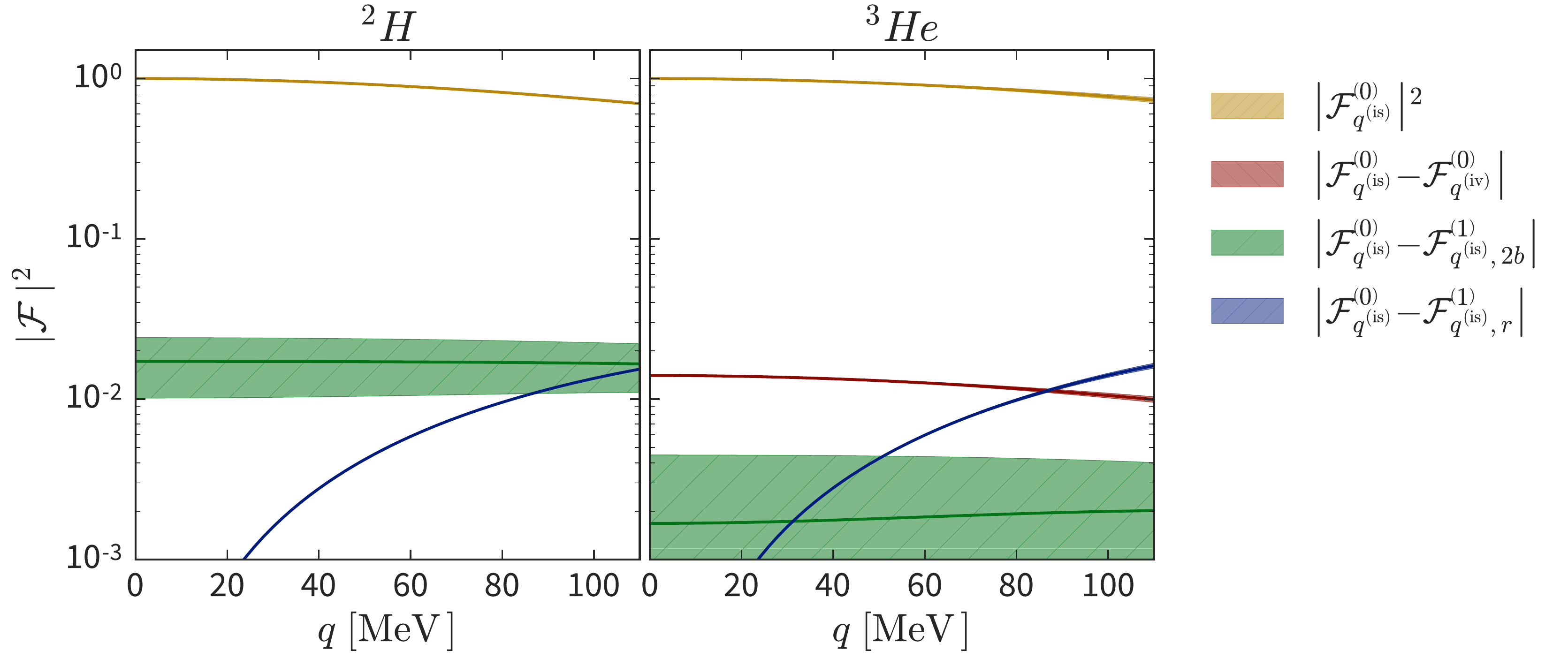}{width=\textwidth}{fig:log_plots_light}{Hierarchy of contributions to $|\mathcal F(\vec q^2)|^2$ for the choice $c_u = c_d$ and $c_s = c_G=0$ using N$^2$LO chiral wave functions  with cut-off $\Lambda_2$.  The diagram displays the absolute values of the various terms.  For more explanation we refer to the main text.}

The pattern is somewhat different for ${}^3$He (and ${}^3$H which is very similar and not shown). In this case the largest correction arises from the interference between the isoscalar and isovector one-nucleon terms. For the choice $c_u =c_d$ this correction amounts to only $(1$-$2)$\% roughly because of the smallness of $\alpha_{q^{\mathrm{(iv)}}}/\alpha_{q^{\mathrm{(is)}}}$, but this can change for different parameter choices. As discussed in the previous section, the two-nucleon contributions are rather small and suffer from $\mathcal O(100\%)$ uncertainties for ${}^3$He in case of N${}^2$LO chiral wave functions.  Despite the large uncertainty, the two-nucleon corrections are smaller than isospin-breaking terms. This is different from scattering off Xe isotopes where two-nucleon corrections were found to be larger than isospin-breaking corrections by roughly an order of magnitude \cite{Hoferichter:2016nvd} for the same choices of $c_u$ and $c_d$. This is not unexpected as the two-nucleon corrections are expected to grow with $A$. Finally, radius corrections are typically smaller than isospin-violating terms.

\subsection{Suppressing the leading-order contributions}\label{sec:cancelation-regime}
In specific scenarios of DM it might occur that specific values of the DM coefficients in Eq.~\eqref{Eq1} lead to cancellations in the leading one-nucleon contributions to the cross sections. In such scenarios, for instance in the framework of isospin-violating DM  \cite{Feng:2011vu}, subleading contributions to the scattering process become relatively enhanced and crucial to include  \cite{Prezeau:2003sv, Cirigliano:2012pq, Cirigliano:2013zta}. In this section we consider a case where isoscalar quark-DM and gluon-DM interactions appear at the same time. We set $c_s=0$ in this section as the strange contributions are almost degenerate with respect to the gluon contributions.

Focusing on isoscalar and gluonic interactions, we see that the differential cross section is proportional to
\begin{equation}
\mathcal F(\vec q^2) = \left(\alpha_{q^{(\mathrm{is})}} + \alpha_{G}\right)   \mathcal{F}^{\left(0\right)}_{q^{(\mathrm{is})}}{\left(\vec q^2\right)} + \alpha_{q^{(\mathrm{is})}}\left(\mathcal{F}^{\left(1\right)}_{q^{(\mathrm{is})},\, r}{\left(\vec q^2\right)}+\mathcal{F}^{\left(1\right)}_{q^{(\mathrm{is})},\, 2b}{\left(\vec q^2\right)}  \right) +  \alpha_{G} \mathcal{F}^{\left(3\right)}_{G,\, 2b} + \dots
\end{equation} 
where we used $\mathcal{F}^{\left(0\right)}_{q^{(\mathrm{is})}}{\left(\vec q^2\right)} = \mathcal{F}^{\left(0\right)}_{G}{\left(\vec q^2\right)}$, see Eq.~\eqref{normalization}. We also included the order (1) isoscalar corrections and the order (3) two-nucleon correction arising from the DM-gluon interactions, see Eq.~\eqref{Glue2body}. We stress that the order (3) currents are not complete and that additional contributions appear at this order. 

The sizes of $\alpha_{q^{(\mathrm{is})}}$ and $\alpha_{G}$ are unknown such that we cannot determine the relative sizes of their contributions in a model-independent way. In this section, we investigate the case where $\alpha_{G}/\alpha_{q^{(\mathrm{is})}} =r$ and set $\alpha_{q^{(\mathrm{is})}} =1$ for simplicity. For values of $\abs{r}\simeq 1$, the order (0) isoscalar and gluon contributions are of similar size. The power counting then predicts that the dominant corrections arise  from the order (1) isoscalar corrections, while the (incomplete) order (3) gluon corrections are expected to be small. As discussed in detail above, for most values of $r$ the leading one-nucleon contributions dominate the scattering process and the subleading terms only provide percent-level corrections. However, around $r=-1$ the leading contributions are suppressed and the subleading terms become relatively enhanced. Clearly, such a scenario corresponds to a tuning of the fundamental couplings $c_u$, $c_d$, and $c_G$, which, admittedly, is rather ad hoc. Nevertheless, similar scenarios have been invoked to reduce the tension between positive DM and negative DM signals in different DM direct detection experiments. Furthermore, the tuned scenarios make higher-order corrections more relevant allowing for a test of the chiral power counting to higher orders. 

We investigate a scenario where a value of $r$ is chosen such that the order (1) currents corrections provide a $50\%$ correction to the differential cross section at $\vec q^2=0$. For the deuteron this turns out to require $r = r_0 \simeq -0.96$. Clearly, the choice for $50\%$ is arbitrary and for values of $-2-r_0 < r < r_0$ the corrections become larger.
In the top panels of Fig.~\ref{fig:fine_tuning_50_mg} we plot $\abs{\mathcal F(\vec q^2)}^2$ for different orders of the current for scattering off ${}^2$H using N${}^2$LO (top-left panel) and N${}^3$LO (top-right panel) chiral wave functions. The green line just includes the order (0) one-nucleon contribution. Because of the cancellation between $\alpha_{q^{(\mathrm{is})}}$ and $\alpha_{G}$, $\mathcal F(\vec q^2)$ is suppressed by $(1+r_0)^2$ with respect to Fig.~\ref{fig:log_plots_light}. The blue line and band also include the order (1) two-nucleon and radius corrections. The uncertainty for N${}^2$LO wave functions is significant, but much reduced once we apply N${}^3$LO wave functions. By construction, the inclusion of the order (1) currents enhances the differential cross section by $50\%$ at zero momentum transfer, but due to the radius corrections this enhancement becomes smaller for larger momentum transfer.

\MyIncFig{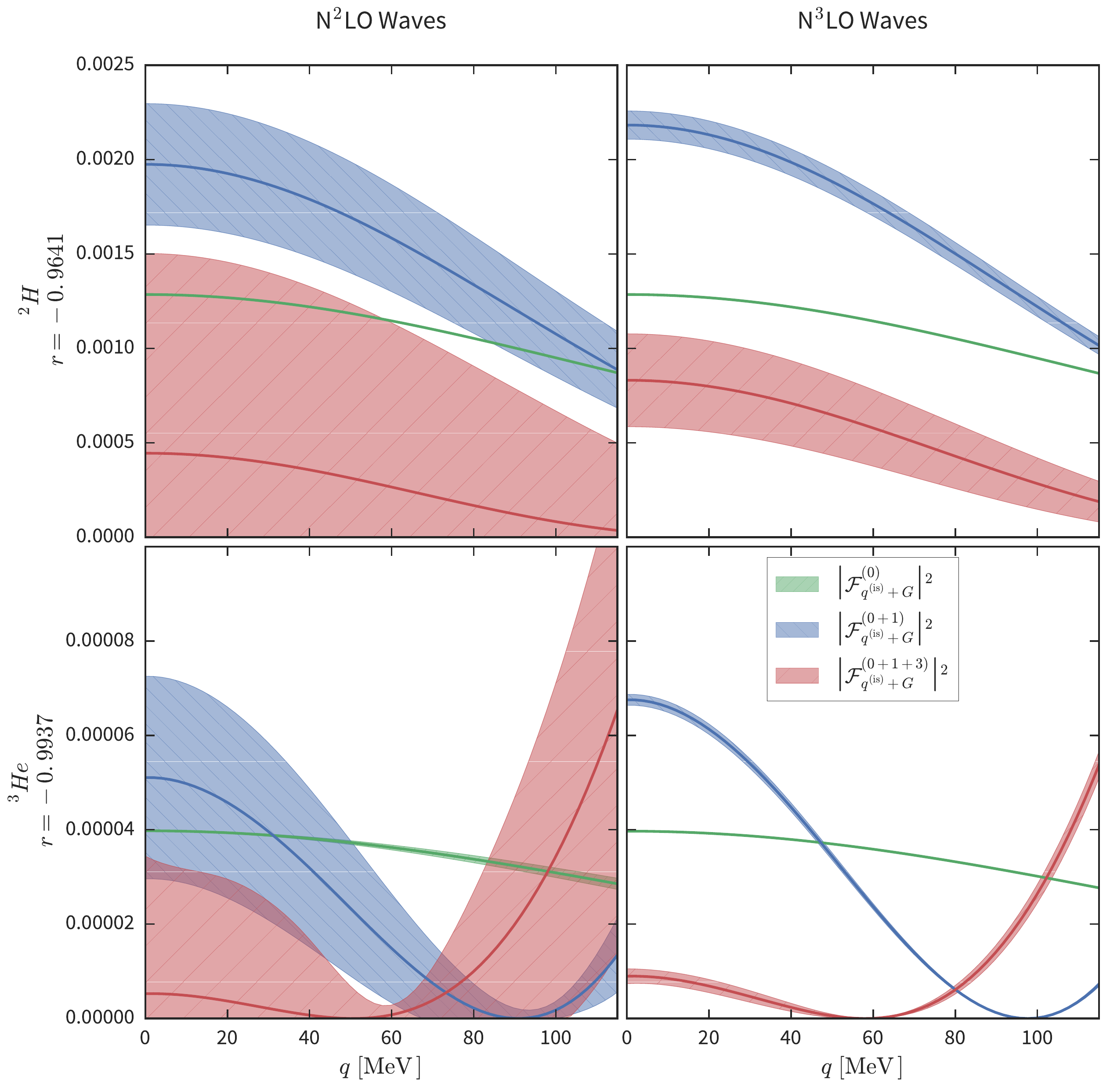}{width=.8\textwidth}{fig:fine_tuning_50_mg}{Plots corresponding to the scenario with nonzero isoscalar and gluonic DM interactions disrussed in Sect.~\ref{sec:cancelation-regime}.  The coefficient $r$ has been chosen such that the order (1) currents provide a $50\%$ correction to the leading-order results.  We plot the square of the response functions $\abs{ \mathcal F }^2 ( \vec q ^2 )$.  The upper (lower) row represents ${}^2$H (${}^3$He) results, while the left (right) panel corresponds to N${}^2$LO (N${}^3$LO) chiral wave functions using cut-off $\Lambda_2$. The green bands correspond to the leading contribution, while the blue and red bands include the order (1) and order (3) currents, respectively.}

We repeat this analysis for scattering off ${}^3$He. Because the order (1) corrections are smaller in this case, we need to further tune $r = r_0 \simeq -0.99$ in order to obtain a $50\%$ correction. In this case, the uncertainties for N${}^2$LO wave functions are large enough to cover the order (0) one-nucleon contributions (the green line), but the results improve significantly once N${}^3$LO wave functions are applied. Radius corrections are relatively more important such that $\mathcal F(\vec q^2)$ vanishes for $| \vec q| \simeq 90$ MeV. 

As discussed in Sect.~\ref{uncertainty}, the uncertainty bands do not cover the effects of missing higher-order currents. To investigate this we also included the contributions from the order (3) two-nucleon correction $\mathcal{F}^{\left(3\right)}_{G,\, 2b}$. The result is depicted by the red line and bands in Fig.~\ref{fig:fine_tuning_50_mg}. Although the power counting indicates that these contributions should be suppressed by $Q^2 \sim m_\pi^2/\Lambda_b^2 \sim 0.05$ with respect to the order (1) currents, they are actually of similar size. In addition, for N${}^2$LO wave functions the uncertainties are large enough to get a result consistent with zero for both ${}^2$H and ${}^3$He. The uncertainty improves once we use N${}^3$LO wave functions, but we still find that adding formally higher-order corrections suppresses $\mathcal F(\vec q^2)$, and thus the differential cross section, by roughly a factor $2.5$ ($7$) for ${}^2$H (${}^3$He). 

The large impact of formally higher-order contributions might seem to indicate that the power counting is not working satisfactory. It must be stressed, however, that we have not included the full order (3) currents, which would consist of many additional terms. In particular, as discussed below Eq.~\eqref{Glue2body}, the order (3) two-nucleon current requires a counter term that appears at the same order for renormalization purposes. Furthermore, it might be that for denser nuclei the chiral series shows better convergence because the order (1) two-nucleon currents could be less suppressed than in light nuclei. Nevertheless, the calculated order (3) contributions can be taken as an order-of-magnitude estimate of the uncertainty due to higher-order corrections to the currents, indicating that scenarios where one-nucleon contributions are suppressed might suffer from sizable additional uncertainties. 

\section{Summary and outlook}\label{Summary}

In this work we have performed calculations of DM scattering off light light nuclei within the framework of chiral perturbation theory. A major aspect is that both ingredients of the calculations, i.e. the nuclear wave functions and the DM-nucleus interactions, are derived from chiral EFT.  In this way, we are able to test the convergence of the chiral series and provide a systematic uncertainty estimate that goes beyond simple cut-off variations. We have focused for now on the lightest bound nuclei: ${}^2$H, ${}^3$H, and ${}^3$He as these systems are simple enough such that the bound-state and scattering equations can be solved with direct methods to very high accuracy and precision. These nuclei are therefore great theoretical tools to study the framework and provide a first step towards calculations on heavier systems with other many-body methods. Furthermore, there has been recent interest in using helium isotopes as DM direct detection targets as light targets are more sensitive to relatively light DM and ${}^3$He can potentially be used for directional DM detection.

We have focused on one particular class of DM interactions at the quark-gluon level, namely scalar interactions that can be parametrized by four parameters describing interactions between light quarks, gluons, and DM particles. At leading order in the chiral expansion each scalar interaction gives rise to an interaction between DM and a single nucleon, while at higher orders also two-nucleon interactions appear. Our framework allows for the systematic inclusion of these, and other, higher-order effects. Two-nucleon interactions are particularly interesting as they can give rise to a different dependence of the DM-nucleus cross section on the atomic number. In fact, we observed that two-nucleon interactions are much larger for ${}^2$H-DM than for ${}^3$He-DM or ${}^3$H-DM scattering, although in both cases the effects are modest. 

We find that the impulse approximation, where it is assumed that the DM-nucleus interactions arise from a sum of DM-nucleon interactions, works well for the scalar DM interactions under investigation. We have provided expressions in terms of Helm form factors fitted to our results that can be used in future studies of DM direct detection using light-nuclear targets. The smallness of two-nucleon contributions for light nuclei, at the few-percent level, justifies the approach of Ref.~\cite{Gazda:2016mrp} at least for scalar DM interactions. The smallness of two-nucleon corrections is  somewhat unexpected as they  appear at next-to-leading order in the chiral expansion indicating a relative contribution at the $(10$-$30)$\% level. The specific spin-isospin properties and/or the diluteness of light nuclei might be the cause for this suppression, and we aim to investigate this in the near future with studies of the deeper-bound ${}^4$He. 

Because two-nucleon corrections are expected to grow for heavier nuclei, as explicitly found in Ref.~\cite{Hoferichter:2016nvd}, we studied how accurate these contributions can be calculated. We found a large dependence on the applied nuclear wave function due to a strong correlation between the D-wave probability of the considered nucleus and the two-nucleon matrix element. 
As such, at least N${}^3$LO chiral nuclear wave functions are required in order to accurately calculate two-nucleon corrections. Whether this is unique to the scalar interactions we investigated here or generalizes to other interactions such as vector or axial-vector currents remains to be seen. We plan to perform such calculations in future work. This is also relevant beyond the topic of DM direct detection as two-nucleon currents are found to be important for, for example, neutrino-nucleus scattering \cite{Lovato:2014eva}. 

Finally, we studied a specific scenario where one-nucleon contributions are suppressed due to a cancellation mechanism such that two-nucleon corrections become crucial to include \cite{Cirigliano:2012pq,Cirigliano:2013zta}. While the systematic nature of our framework allows for controlled calculations of such scenarios, we found that the missing higher-order currents can become relevant. However, as the currents at higher orders have not been fully developed a more careful analysis has to be postponed. 

As becomes clear from the above discussion, much work still remains to be done. Our calculations need to be generalized to other DM interactions along the lines of Ref.~\cite{Hoferichter:2015ipa}. Furthermore, we need to extend the calculations to heavier nuclei. As the four-nucleon system can be solved with methods analogous to those applied here, our first target is the ${}^4$He nucleus. This system is interesting as it is a candidate for a direct-detection target \cite{Guo:2013dt} and is much denser then nuclei considered in this work. More specific, we want to study the correlation between the D-wave probability and isoscalar two-nucleon matrix elements. In recent years major progress has been made in first-principle calculations of light- to medium-heavy nuclei. It would be extremely interesting to use methods such as nuclear lattice EFT \cite{Lahde:2013uqa}, the Jacobi no-core shell model \cite{Liebig:2015kwa}, and quantum Monte Carlo methods \cite{Carlson:2014vla} to perform systematic calculations on heavier systems for the fascinating problem of DM direct detection.

\section*{Acknowledgements}
 JdV  acknowledges  support by the Dutch Organization for Scientific Research (NWO) 
through a VENI grant.  We are very grateful to Martin Hoferichter, Evgeny Epelbaum, Hermann Krebs, Ulf Mei{\ss}ner, and Thomas Luu for valuable discussions and comments on the manuscript.  The numerical calculations have been performed on JUQUEEN and JURECA of the J\"ulich Supercomputing Centre, J\"ulich, Germany.

\appendix
\section{Operator expectation values}
\label{appendix:2pi_res}
The computation of the spin-independent scattering matrix in Eq.~\eqref{eq:current_to_scattering_matrix} involves the sum (average) over all outgoing (incoming) nuclear polarizations.  This summation can be simplified in terms of reduced matrix elements
\begin{equation}
	\braket{ j' m_j' | \hat{\mathcal{O}} _ {\xi m_\xi} | j m_j  }
	=
	\CG{j}{\xi}{j'}{m_j}{m_\xi}{m_j'}
	\braket{j' | \hat{\mathcal{O}} _ {\xi} | j}
	\ .
\end{equation}
Eq.~\eqref{eq:current_to_scattering_matrix} then becomes
\begin{equation}
	\abs{ \mathcal{M}_A ( \vec q ^2) }^2 = 
	(2 m_T )^2( 2 m_\chi )^2
	\sum_{\xi=0}^\infty
	\abs{
		\Braket{ \Psi_T, j | \ \hat{ J } (\vec q ^2, \xi) \ | \Psi_T, j }
	}^2 \ .
\end{equation}
In the limit of $q \rightarrow 0$, one can show that for spin-conserving structures, only contributions with $\xi=0$ survive.  Following Eq.~\eqref{eq:def-response_functions}, we further define
\begin{equation}
	\label{def:op_expectation_value}
	\braket{ \hat J_a }_T
	:=
	\frac{
		\Braket{ \Psi_T, j | \ \hat{ J }_{ a} (\vec q ^2, \xi) \ | \Psi_T, j }
	}{A \, \abs{ \sigma_{\pi N} \ c_{\chi} } } \Bigg|_{\xi=0}
	\ .
\end{equation}
For benchmarking purposes, we present the expectation values for the isoscalar quark-DM two-nucleon contributions at $\vec q=0$  in Table \ref{tab:amp_res}.  We estimate the numerical uncertainty to be below the  $1 \%$-level with respect to the value of each matrix element.


\begin{table}
\scriptsize
\centering
\begin{tabular}{l | rrr}
NN interaction &     $\braket{ \hat J_{q^{(\mathrm{is})},2b} }_{^2\mathrm{H}}$ &     $\braket{\hat J_{q^{(\mathrm{is})},2b} }_{^3\mathrm{H}}$ &    $\braket{\hat J_{q^{(\mathrm{is})},2b} }_{^3\mathrm{He}}$ \\[0.4em]
\hline\hline
AV18 + Urb IXF          &         / &  2.62$\cdot 10^{-3}$ &  2.52$\cdot 10^{-3}$ \\[-0.4em]
AV18                    &  1.58$\cdot 10^{-4}$ &  2.62$\cdot 10^{-3}$ &  2.83$\cdot 10^{-3}$ \\[-0.4em]
CD-Bonn + TM            &         / &  1.38$\cdot 10^{-3}$ &  1.51$\cdot 10^{-3}$ \\[-0.4em]
CD-Bonn                 &  1.17$\cdot 10^{-4}$ &  1.47$\cdot 10^{-3}$ &  1.62$\cdot 10^{-3}$ \\[-0.4em]
NIJM                    &  1.57$\cdot 10^{-4}$ &         / &         / \\
\hline
LO\,\,\,\,\,\, $(Q^0)$   $\Lambda_{1}$ &  2.28$\cdot 10^{-3}$ & -1.46$\cdot 10^{-3}$ & -1.45$\cdot 10^{-3}$ \\[-0.4em]
LO\,\,\,\,\,\, $(Q^0)$   $\Lambda_{2}$ & -5.95$\cdot 10^{-4}$ & -1.77$\cdot 10^{-3}$ & -1.77$\cdot 10^{-3}$ \\[-0.4em]
LO\,\,\,\,\,\, $(Q^0)$   $\Lambda_{3}$ & -2.66$\cdot 10^{-3}$ & -2.01$\cdot 10^{-3}$ & -2.02$\cdot 10^{-3}$ \\[-0.4em]
LO\,\,\,\,\,\, $(Q^0)$   $\Lambda_{4}$ & -4.12$\cdot 10^{-3}$ & -2.20$\cdot 10^{-3}$ & -2.22$\cdot 10^{-3}$ \\[-0.4em]
LO\,\,\,\,\,\, $(Q^0)$   $\Lambda_{5}$ & -5.13$\cdot 10^{-3}$ & -2.36$\cdot 10^{-3}$ & -2.38$\cdot 10^{-3}$ \\
\hline 
NLO\,\, ($Q^{2}$)    $\Lambda_{1}$ &  1.17$\cdot 10^{-4}$ &  1.48$\cdot 10^{-3}$ &  1.62$\cdot 10^{-3}$ \\[-0.4em]
NLO\,\, ($Q^{2}$)    $\Lambda_{2}$ &  8.80$\cdot 10^{-3}$ &  7.08$\cdot 10^{-4}$ &  8.18$\cdot 10^{-4}$ \\[-0.4em]
NLO\,\, ($Q^{2}$)    $\Lambda_{3}$ &  6.09$\cdot 10^{-3}$ &  2.72$\cdot 10^{-5}$ &  1.08$\cdot 10^{-4}$ \\[-0.4em]
NLO\,\, ($Q^{2}$)    $\Lambda_{4}$ &  3.70$\cdot 10^{-3}$ & -5.47$\cdot 10^{-4}$ & -4.91$\cdot 10^{-4}$ \\[-0.4em]
NLO\,\, ($Q^{2}$)    $\Lambda_{5}$ &  1.65$\cdot 10^{-3}$ & -1.01$\cdot 10^{-3}$ & -9.78$\cdot 10^{-4}$ \\
\hline 
N${}^2$LO ($Q^{3}$)   $\Lambda_{1}$ &  1.10$\cdot 10^{-4}$ &  1.37$\cdot 10^{-3}$ &  1.51$\cdot 10^{-3}$ \\[-0.4em]
N${}^2$LO ($Q^{3}$)   $\Lambda_{2}$ &  8.59$\cdot 10^{-3}$ &  7.25$\cdot 10^{-4}$ &  8.39$\cdot 10^{-4}$ \\[-0.4em]
N${}^2$LO ($Q^{3}$)   $\Lambda_{3}$ &  6.17$\cdot 10^{-3}$ &  1.06$\cdot 10^{-4}$ &  1.93$\cdot 10^{-4}$ \\[-0.4em]
N${}^2$LO ($Q^{3}$)   $\Lambda_{4}$ &  3.97$\cdot 10^{-3}$ & -4.39$\cdot 10^{-4}$ & -3.75$\cdot 10^{-4}$ \\[-0.4em]
N${}^2$LO ($Q^{3}$)   $\Lambda_{5}$ &  2.04$\cdot 10^{-3}$ & -8.98$\cdot 10^{-4}$ & -8.55$\cdot 10^{-4}$ \\
\hline 
N${}^3$LO ($Q^{4}$)   $\Lambda_{1}$ &  7.31$\cdot 10^{-3}$ &  7.54$\cdot 10^{-4}$ &  8.61$\cdot 10^{-4}$ \\[-0.4em]
N${}^3$LO ($Q^{4}$)   $\Lambda_{2}$ &  1.09$\cdot 10^{-4}$ &  1.66$\cdot 10^{-3}$ &  1.81$\cdot 10^{-3}$ \\[-0.4em]
N${}^3$LO ($Q^{4}$)   $\Lambda_{3}$ &  1.29$\cdot 10^{-4}$ &  2.21$\cdot 10^{-3}$ &  2.38$\cdot 10^{-3}$ \\[-0.4em]
N${}^3$LO ($Q^{4}$)   $\Lambda_{4}$ &  1.37$\cdot 10^{-4}$ &  2.44$\cdot 10^{-3}$ &  2.62$\cdot 10^{-3}$ \\[-0.4em]
N${}^3$LO ($Q^{4}$)   $\Lambda_{5}$ &  1.38$\cdot 10^{-4}$ &  2.50$\cdot 10^{-3}$ &  2.69$\cdot 10^{-3}$ \\
\hline 
N${}^4$LO ($Q^{5}$)   $\Lambda_{1}$ &  1.06$\cdot 10^{-4}$ &  1.62$\cdot 10^{-3}$ &  1.77$\cdot 10^{-3}$ \\[-0.4em]
N${}^4$LO ($Q^{5}$)   $\Lambda_{2}$ &  1.13$\cdot 10^{-4}$ &  1.77$\cdot 10^{-3}$ &  1.92$\cdot 10^{-3}$ \\[-0.4em]
N${}^4$LO ($Q^{5}$)   $\Lambda_{3}$ &  1.18$\cdot 10^{-4}$ &  1.91$\cdot 10^{-3}$ &  2.07$\cdot 10^{-3}$ \\[-0.4em]
N${}^4$LO ($Q^{5}$)   $\Lambda_{4}$ &  1.25$\cdot 10^{-4}$ &  2.10$\cdot 10^{-3}$ &  2.27$\cdot 10^{-3}$ \\[-0.4em]
N${}^4$LO ($Q^{5}$)   $\Lambda_{5}$ &  1.27$\cdot 10^{-4}$ &  2.20$\cdot 10^{-3}$ &  2.38$\cdot 10^{-3}$ \\
\end{tabular}
\caption{\label{tab:amp_res}%
Central values for spin-independent isoscalar-quark two-pion-exchange operators $\hat J_{q^{(\mathrm{is})},2b}$ for all applied wave functions at zero momentum transfer $\vec q=0$. Fore more details, we refer to Appendix \ref{appendix:2pi_res}.%
}
\end{table}

\bibliographystyle{h-physrev3} 
\bibliography{bibliography}

\begin{thebibliography}{10}

\bibitem{Bertone:2016nfn}
G.~Bertone and D.~Hooper,
\newblock Submitted to: Rev. Mod. Phys.  (2016), 1605.04909.

\bibitem{Akerib:2016vxi}
LUX, D.~S. Akerib {\em et~al.},
\newblock Phys. Rev. Lett. {\bf 118}, 021303 (2017), 1608.07648.

\bibitem{Fu:2016ega}
PandaX-II, C.~Fu {\em et~al.},
\newblock Phys. Rev. Lett. {\bf 118}, 071301 (2017), 1611.06553.

\bibitem{Aprile:2015uzo}
XENON, E.~Aprile {\em et~al.},
\newblock JCAP {\bf 1604}, 027 (2016), 1512.07501.

\bibitem{Akerib:2015cja}
LZ, D.~S. Akerib {\em et~al.},
\newblock (2015), 1509.02910.

\bibitem{Fan:2010gt}
J.~Fan, M.~Reece, and L.-T. Wang,
\newblock JCAP {\bf 1011}, 042 (2010), 1008.1591.

\bibitem{Fitzpatrick:2012ix}
A.~L. Fitzpatrick, W.~Haxton, E.~Katz, N.~Lubbers, and Y.~Xu,
\newblock JCAP {\bf 1302}, 004 (2013), 1203.3542.

\bibitem{Cirigliano:2012pq}
V.~Cirigliano, M.~L. Graesser, and G.~Ovanesyan,
\newblock JHEP {\bf 10}, 025 (2012), 1205.2695.

\bibitem{Klos:2013rwa}
P.~Klos, J.~Menendez, D.~Gazit, and A.~Schwenk,
\newblock Phys. Rev. {\bf D88}, 083516 (2013), 1304.7684,
\newblock [Erratum: Phys. Rev.D89,no.2,029901(2014)].

\bibitem{Hill:2014yxa}
R.~J. Hill and M.~P. Solon,
\newblock Phys. Rev. {\bf D91}, 043505 (2015), 1409.8290.

\bibitem{Hoferichter:2015ipa}
M.~Hoferichter, P.~Klos, and A.~Schwenk,
\newblock Phys. Lett. {\bf B746}, 410 (2015), 1503.04811.

\bibitem{Bishara:2016hek}
F.~Bishara, J.~Brod, B.~Grinstein, and J.~Zupan,
\newblock JCAP {\bf 02}, 009 (2017), 1611.00368.

\bibitem{Prezeau:2003sv}
G.~Prezeau, A.~Kurylov, M.~Kamionkowski, and P.~Vogel,
\newblock Phys. Rev. Lett. {\bf 91}, 231301 (2003), astro-ph/0309115.

\bibitem{Hoferichter:2016nvd}
M.~Hoferichter, P.~Klos, J.~Menendez, and A.~Schwenk,
\newblock Phys. Rev. {\bf D94}, 063505 (2016), 1605.08043.

\bibitem{Beane:2013kca}
S.~R. Beane, S.~D. Cohen, W.~Detmold, H.~W. Lin, and M.~J. Savage,
\newblock Phys. Rev. {\bf D89}, 074505 (2014), 1306.6939.

\bibitem{Cirigliano:2013zta}
V.~Cirigliano, M.~L. Graesser, G.~Ovanesyan, and I.~M. Shoemaker,
\newblock Phys. Lett. {\bf B739}, 293 (2014), 1311.5886.

\bibitem{Guo:2013dt}
W.~Guo and D.~N. McKinsey,
\newblock Phys. Rev. {\bf D87}, 115001 (2013), 1302.0534.

\bibitem{SNONEWS}
http://www.snolab.ca/news projects/NEWSSNO.html.

\bibitem{Schutz:2016tid}
K.~Schutz and K.~M. Zurek,
\newblock Phys. Rev. Lett. {\bf 117}, 121302 (2016), 1604.08206.

\bibitem{Profumo:2015oya}
S.~Profumo,
\newblock Phys. Rev. {\bf D93}, 055036 (2016), 1507.07531.

\bibitem{Franarin:2016ppr}
T.~Franarin and M.~Fairbairn,
\newblock Phys. Rev. {\bf D94}, 053004 (2016), 1605.08727.

\bibitem{Gazda:2016mrp}
D.~Gazda, R.~Catena, and C.~Forssen,
\newblock (2016), 1612.09165.

\bibitem{Carlson:2014vla}
J.~Carlson {\em et~al.},
\newblock Rev. Mod. Phys. {\bf 87}, 1067 (2015), 1412.3081.

\bibitem{Lahde:2013uqa}
T.~A. L{\"a}hde {\em et~al.},
\newblock Phys. Lett. {\bf B732}, 110 (2014), 1311.0477.

\bibitem{Shifman:1978zn}
M.~A. Shifman, A.~I. Vainshtein, and V.~I. Zakharov,
\newblock Phys. Lett. {\bf B78}, 443 (1978).

\bibitem{Epelbaum:2014sza}
E.~Epelbaum, H.~Krebs, and U.-G. Mei{\ss}ner,
\newblock Phys. Rev. Lett. {\bf 115}, 122301 (2015), 1412.4623.

\bibitem{Weinberg:1978kz}
S.~Weinberg,
\newblock Physica {\bf A96}, 327 (1979).

\bibitem{Gasser:1983yg}
J.~Gasser and H.~Leutwyler,
\newblock Annals Phys. {\bf 158}, 142 (1984).

\bibitem{Weinberg:1990rz}
S.~Weinberg,
\newblock Phys. Lett. {\bf B251}, 288 (1990).

\bibitem{Ordonez:1993tn}
C.~Ordonez, L.~Ray, and U.~van Kolck,
\newblock Phys. Rev. Lett. {\bf 72}, 1982 (1994).

\bibitem{Bernard:1995dp}
V.~Bernard, N.~Kaiser, and U.-G. Mei{\ss}ner,
\newblock Int.J.Mod.Phys. {\bf E4}, 193 (1995), hep-ph/9501384.

\bibitem{Hoferichter:2015tha}
M.~Hoferichter, J.~Ruiz~de Elvira, B.~Kubis, and U.-G. Mei{\ss}ner,
\newblock Phys. Rev. Lett. {\bf 115}, 192301 (2015), 1507.07552.

\bibitem{Agashe:2014kda}
Particle Data Group, K.~Olive {\em et~al.},
\newblock Chin.Phys. {\bf C38}, 090001 (2014).

\bibitem{Crivellin:2013ipa}
A.~Crivellin, M.~Hoferichter, and M.~Procura,
\newblock Phys. Rev. {\bf D89}, 054021 (2014), 1312.4951.

\bibitem{Epelbaum:2014efa}
E.~Epelbaum, H.~Krebs, and U.-G. Mei{\ss}ner,
\newblock Eur. Phys. J. A {\bf 51}, 53 (2015).

\bibitem{Crivellin:2015bva}
A.~Crivellin, M.~Hoferichter, M.~Procura, and L.~C. Tunstall,
\newblock JHEP {\bf 07}, 129 (2015), 1503.03478.

\bibitem{Hoferichter:2015dsa}
M.~Hoferichter, J.~Ruiz~de Elvira, B.~Kubis, and U.-G. Mei{\ss}ner,
\newblock Phys. Rev. Lett. {\bf 115}, 092301 (2015), 1506.04142.

\bibitem{Brantley:2016our}
D.~A. Brantley {\em et~al.},
\newblock (2016), 1612.07733.

\bibitem{Durr:2015dna}
S.~Durr {\em et~al.},
\newblock Phys. Rev. Lett. {\bf 116}, 172001 (2016), 1510.08013.

\bibitem{Hoferichter:2012wf}
M.~Hoferichter, C.~Ditsche, B.~Kubis, and U.-G. Mei{\ss}ner,
\newblock JHEP {\bf 06}, 063 (2012), 1204.6251.

\bibitem{Martinprivate}
M.~Hoferichter,
\newblock Private communication  (2017).

\bibitem{Fettes:2000gb}
N.~Fettes, U.-G. Mei{\ss}ner, M.~Mojzis, and S.~Steininger,
\newblock Annals Phys. {\bf 283}, 273 (2000), hep-ph/0001308,
\newblock [Erratum: Annals Phys.288,249(2001)].

\bibitem{Junnarkar:2013ac}
P.~Junnarkar and A.~Walker-Loud,
\newblock Phys.Rev. {\bf D87}, 114510 (2013), 1301.1114.

\bibitem{DelNobile:2013sia}
M.~Cirelli, E.~Del~Nobile, and P.~Panci,
\newblock JCAP {\bf 1310}, 019 (2013), 1307.5955.

\bibitem{Smith:2006ym}
M.~C. Smith {\em et~al.},
\newblock Mon. Not. Roy. Astron. Soc. {\bf 379}, 755 (2007), astro-ph/0611671.

\bibitem{Wiringa:1995co}
R.~B. Wiringa, V.~G.~J. Stoks, and R.~Schiavilla,
\newblock Phys. Rev. {\bf C51}, 38 (1995), nucl-th/9408016.

\bibitem{Machleidt:2001ib}
R.~Machleidt,
\newblock Phys. Rev. {\bf C 63}, 024001 (2001).

\bibitem{VanDerLeun:1982gh}
C.~Van Der~Leun and C.~Alderliesten,
\newblock Nucl. Phys. A {\bf 380}, 261 (1982).

\bibitem{Huber:1998fp}
A.~Huber {\em et~al.},
\newblock Phys. Rev. Lett. {\bf 80}, 468 (1998).

\bibitem{Bishop:1979zz}
D.~M. Bishop and L.~M. Cheung,
\newblock Phys. Rev. {\bf A20}, 381 (1979).

\bibitem{Borbely:1985tur}
I.~Borbely, W.~Gr{\"u}ebler, V.~Konig, P.~A. Schmelzbach, and A.~M.
  Mukhamedzhanov,
\newblock Phys. Lett. {\bf B160}, 17 (1985).

\bibitem{Rodning:1990hm}
N.~L. Rodning and L.~D. Knutson,
\newblock Phys. Rev. C {\bf 41}, 898 (1990).

\bibitem{Weinberg:1992jd}
S.~Weinberg,
\newblock Phys. Lett. {\bf B295}, 114 (1992).

\bibitem{Birse:2006jn}
M.~C. Birse,
\newblock Phys. Rev. {\bf C 74}, 014003 (2006).

\bibitem{Nogga:2005hb}
A.~Nogga, R.~G.~E. Timmermans, and U.~van Kolck,
\newblock Phys. Rev. {\bf C 72}, 054006 (2005).

\bibitem{PavonValderrama:2006gp}
M.~Pav{\'o}n~Valderrama and E.~Ruiz~Arriola,
\newblock Phys. Rev. {\bf C 74}, 054001 (2006).

\bibitem{PavonValderrama:2005uj}
M.~Pav{\'o}n~Valderrama and E.~Ruiz~Arriola,
\newblock Phys. Rev. C {\bf 74}, 064004 (2006).

\bibitem{Epelbaum:2006pt}
E.~Epelbaum and U.-G. Mei{\ss}ner,
\newblock Few Body Syst. {\bf 54}, 2175 (2013), nucl-th/0609037.

\bibitem{Valderrama:2011hw}
M.~Pav{\'o}n~Valderrama,
\newblock Phys. Rev. {\bf C 84}, 064002 (2011).

\bibitem{Valderrama:2011hz}
M.~Pav{\'o}n~Valderrama,
\newblock Phys. Rev. {\bf C 83}, 024003 (2011).

\bibitem{Epelbaum:2009hn}
E.~Epelbaum and J.~Gegelia,
\newblock Eur. Phys. J. {\bf A41}, 341 (2009), 0906.3822.

\bibitem{Binder:2016bi}
S.~Binder {\em et~al.},
\newblock Phys. Rev. C {\bf nucl-th}, 044002 (2015).

\bibitem{Nogga:2003iy}
A.~Nogga {\em et~al.},
\newblock Phys. Rev. {\bf C 67}, 034004 (2003).

\bibitem{Nuclearandhypernuc:2001wd}
A.~Nogga,
\newblock {\em {Nuclear and hypernuclear three- and four-body bound states}},
\newblock PhD thesis, Bochum University, 2001.

\bibitem{Audi:2012dv}
G.~Audi {\em et~al.},
\newblock Chinese Phys. C {\bf 36}, 1287 (2012).

\bibitem{Wang:2012fh}
M.~Wang {\em et~al.},
\newblock Chinese Phys. C {\bf 36}, 1603 (2012).

\bibitem{Nogga:2002kw}
A.~Nogga, H.~Kamada, W.~Gl{\"o}ckle, and B.~R. Barrett,
\newblock Phys. Rev. {\bf C 65}, 054003 (2002).

\bibitem{Pudliner:1997fv}
B.~S. Pudliner, V.~R. Pandharipande, J.~Carlson, S.~C. Pieper, and R.~B.
  Wiringa,
\newblock Phys. Rev. {\bf C56}, 1720 (1997), nucl-th/9705009.

\bibitem{Coon:2001it}
S.~A. Coon and H.~K. Han,
\newblock Few-Body Syst {\bf 30}, 131 (2001).

\bibitem{Friar:1987hr}
J.~L. Friar, B.~F. Gibson, and G.~L. Payne,
\newblock Phys. Rev. {\bf C 35}, 1502 (1987).

\bibitem{Ottermann:1985km}
C.~R. Ottermann {\em et~al.},
\newblock Nucl. Phys. {\bf A436}, 688 (1985).

\bibitem{Baroni:2015uza}
A.~Baroni, L.~Girlanda, S.~Pastore, R.~Schiavilla, and M.~Viviani,
\newblock Phys. Rev. {\bf C93}, 015501 (2016), 1509.07039,
\newblock [Erratum: Phys. Rev.C93,no.4,049902(2016)].

\bibitem{Krebs:2016rqz}
H.~Krebs, E.~Epelbaum, and U.-G. Mei{\ss}ner,
\newblock Annals Phys. {\bf 378}, 317 (2017), 1610.03569.

\bibitem{McCabe:2010zh}
C.~McCabe,
\newblock Phys. Rev. {\bf D82}, 023530 (2010), 1005.0579.

\bibitem{Liebig:2010ki}
S.~Liebig, V.~Baru, F.~Ballout, C.~Hanhart, and A.~Nogga,
\newblock Eur. Phys. J. {\bf A47}, 69 (2011), 1003.3826.

\bibitem{Baru:2011bw}
V.~Baru {\em et~al.},
\newblock Nucl. Phys. {\bf A872}, 69 (2011), 1107.5509.

\bibitem{HelmFF}
R.~H. Helm,
\newblock Phys. Rev. {\bf 104}, 1466 (1956).

\bibitem{Lewin1996}
J.~D. Lewin and P.~F. Smith,
\newblock Astroparticle Physics {\bf 6}, 87 (1996).

\bibitem{Lovato:2014eva}
A.~Lovato, S.~Gandolfi, J.~Carlson, S.~C. Pieper, and R.~Schiavilla,
\newblock Phys. Rev. Lett. {\bf 112}, 182502 (2014), 1401.2605.

\bibitem{Feng:2011vu}
J.~L. Feng, J.~Kumar, D.~Marfatia, and D.~Sanford,
\newblock Phys. Lett. {\bf B703}, 124 (2011), 1102.4331.

\bibitem{Liebig:2015kwa}
S.~Liebig, U.-G. Mei{\ss}ner, and A.~Nogga,
\newblock Eur. Phys. J. {\bf A52}, 103 (2016), 1510.06070.

\end{thebibliography}

\batchmode
\end{document}